\documentclass{article}

\usepackage[english]{babel}

\usepackage[letterpaper,top=2cm,bottom=2cm,left=3cm,right=3cm,marginparwidth=1.75cm]{geometry}

\usepackage[backend=bibtex, style=numeric, sorting=none]{biblatex}
\usepackage{amsmath}
\usepackage{graphicx}
\usepackage[colorlinks=true, allcolors=blue]{hyperref}
\usepackage{graphicx}
\usepackage{float}
\usepackage{titlesec}
\usepackage{multirow}
\usepackage{longtable}
\usepackage{lscape}
\usepackage[width=\textwidth]{caption}
\usepackage{changepage}
\usepackage{authblk}
\usepackage{hyperref}

\providecommand{\keywords}[1]
{
  \small	
  \textbf{\textit{Keywords---}} #1
}

\title{Machine Learning Small Molecule Properties in Drug Discovery}
\author[a,b]{Nikolai Schapin\thanks{corresponding author, e-mail address: n.shapin@acellera.com}}
\author[a]{Maciej Majewski}
\author[a]{Alejandro Varela}
\author[a]{Carlos Arroniz}
\author[b,c,d]{Gianni De Fabritiis\thanks{corresponding author, e-mail address: g.defabritiis@acellera.com}}
\affil[a]{Acellera Labs, C/ Doctor Trueta 183, 08005 Barcelona, Spain}
\affil[b]{Computational Science Laboratory, Universitat Pompeu Fabra, PRBB, C/ Doctor Aiguader 88, 08003 Barcelona, Spain}
\affil[c]{Institució Catalana de Recerca i Estudis Avançats (ICREA), Passeig Lluís Companys 23, 08010 Barcelona, Spain}
\affil[d]{Acellera, Devonshire House 582 Honeypot Lane Stanmore Middlesex,HA7 1JS United Kingdom}

\begin{document}
\maketitle

\begin{abstract}
Machine learning (ML) is a promising approach for predicting small molecule properties in drug discovery. Here, we provide a comprehensive overview of various ML methods introduced for this purpose in recent years. We review a wide range of properties, including binding affinities, solubility, and ADMET (Absorption, Distribution, Metabolism, Excretion, and Toxicity). We discuss existing popular datasets and molecular descriptors and embeddings, such as chemical fingerprints and graph-based neural networks. We highlight also challenges of predicting and optimizing multiple properties during hit-to-lead and lead optimization stages of drug discovery and explore briefly possible multi-objective optimization techniques that can be used to balance diverse properties while optimizing lead candidates. Finally,  techniques to provide an understanding of model predictions, especially for critical decision-making in drug discovery are assessed.
Overall, this review provides insights into the landscape of ML models for small molecule property predictions in drug discovery. So far, there are multiple diverse approaches, but their performances are often comparable. Neural networks, while more flexible, do not always outperform simpler models. This shows that the availability of high-quality training data remains crucial for training accurate models and there is a need for standardized benchmarks, additional performance metrics, and best practices to enable richer comparisons between the different techniques and models that can shed a better light on the differences between the many techniques.
\end{abstract}


\section{Introduction}

Early stage, preclinical drug discovery is a step-wise process, where at each stage, hit molecules are required to meet certain criteria to ensure their efficacy and quality before proceeding to the next stage. This results in a series of molecular properties that need to be optimized. In order to do this, they need to be measured, which traditionally is done through wet-lab experiments that are costly and time-consuming. The  estimated R\&D expenditure is around 41 billion euros in Europe and 83 billion US dollars in the USA with R\&D costs per drug ranging around 1-2 billion US dollars \cite {cost_randd1,cost_randd2}. This process takes on average 10-13 years \cite{cost_randd1,cost_randd2}, and only 1 out of 10 000 substances tested \cite{cost_randd2} will pass through all the stages to become a new successfully marketed drug. While most of the cost and two-thirds of the time are linked to the stages of clinical trials, most of the candidate molecules  fail during these stages \cite{attrition_clinical_stages}. The main reasons \cite{reasons_failure_drugs} for failure are low efficacy of the drug, high toxicity, or commercial reasons. The first two are often the result of unsuccessful or insufficient establishment of key molecular properties of the hit and lead candidates during the early stages of drug discovery.

Various resource-efficient computational techniques have been developed which all fall under the group of computer-aided drug design methods \cite{cadd} in order to improve the initial screening of compounds in these early stages by both increasing the amount of screened compounds and enabling compound selection and prioritization. These various methods use computational algorithms and models to estimate the molecular properties of screened compounds without having to recur to expensive laboratory experimentation. 

Various groups of methods have been developed such as docking algorithms, molecular dynamics algorithms, quantum mechanical/molecular mechanical (QM/MM) simulations and empirical scoring functions. 
Docking algorithms \cite{docking} provide a way to compute binding poses of ligands to their targets by using interaction information between both and generating a score. This score can also be used to approximate binding affinities, however, it has been shown to produce low accuracy results \cite{docking_failure,docking_bad}. 

More accurate estimates of binding affinity can be obtained with molecular dynamics or QM/MM simulations \cite{md} that simulate atom and molecular movements using computations to approximate the force field that drives this atomic movement. Binding affinities can then be estimated from the forces acting on the molecular structures. A drawback of these methods is that they are computationally expensive, making it difficult to analyze large amounts of compounds. A faster approach is the use of empirical scoring functions \cite{empirical_models}. This approach involves the application of molecular or structural descriptors of screened compounds, such as quantitative structure-activity relationship (QSAR) or quantitative structure-property relationship (QSPR) models, or representations of protein-ligand complexes together with either fixed functional form models or more complex machine learning models to predict different molecular properties. Fixed functional form models can lack accuracy as they use manually derived rules which can fail to model complex relationships found in the input data. Machine learning models can improve on this by using more flexible models that can model both linear and non-linear relationships. Furthermore, as they learn strictly from the observations, human construction bias can be avoided. Various machine learning methods \cite{ml_dd} have therefore been developed for different tasks within the drug discovery process. 

In this review paper, we will take a look at the current role of machine learning in predicting specifically molecular properties of small molecules for drug discovery. We will analyze how machine learning techniques have been applied so far to determine various important molecular properties. We will assess whether they were able to reach enough accuracy and speed and whether they found a practical implementation in drug discovery. We will also comment on the potential disadvantages of these techniques and their causes. Lastly, we will formulate conclusions on the current state of these techniques in drug discovery and suggest some interesting areas for future research in this field.

\section{Model Types}\label{sec:model_types}

Different types of machine learning models with varying degrees of complexity can predict molecular properties such as similarity-based models, linear models, kernel-based models, Bayesian models, tree-based models, and neural networks. 

Similarity-based models try to generate predictions for provided compounds based on known molecular properties of similar compounds. Linear models use linear functional forms to learn linear relationships between representations of input molecular structures and their molecular properties. Kernel-based models transform the input data using kernel functions in order to improve the predictions of the data. Bayesian models use the logic of Bayesian inference where they fit a function of loose form to the data in order to minimize the difference between the obtained likelihood distribution of model predictions and the true posterior distribution of the data. Tree-based models make use of single or multiple decision trees constructed from a selection of features from the input structures. Lastly, neural networks employ models that consist of many interconnected neurons organized in layers that perform non-linear transformations and aggregations of the input data to learn complex, non-linear relationships. A more detailed description and examples for each type are discussed and presented in the following subsections.

\subsection{Similarity-based Models}

The first group describes a simple and straightforward technique which is the k-nearest neighbors (kNN) technique \cite{knn}. This method is based on similarities between the data points and uses the principle that similar data points will also yield a similar outcome value. It performs predictions about new data points by applying a weighted average between the k-nearest neighbors of that data point where k can be a user-defined integer value. One can use different similarity measures to compute the similarities. Closely related, the nearest centroid method \cite{nearest_centroid} uses the closest distance to the center of the established clusters instead. As the latter method uses distances to cluster center points rather than individual most similar data points, it is more suited for classification tasks rather than regression.

\subsection{Linear Models}

Different types of linear models exist which mainly differ in the ways how they fit the linear functions to the input data.
Linear models and multivariate linear models in the case of multiple variables, work by fitting a linear function to the data in order to either establish a regression function like multivariate linear regression (MLR) or a linear classification boundary. The data itself is presented as a feature vector and the resulting fitted linear function will have the same dimension as the input data feature vector.  Flexible discriminant analysis \cite{flex_discrim_analysis} can be seen as an extension of linear models for multi-class classification problems. It uses non-parametric regression as opposed to classic linear discriminant analysis to find groups of data and applies further classic linear discriminant analysis to maximize the separation of the data between the found groups. Multivariate adaptive regression splines (MARS) \cite{mars} is another extension of linear models where the full dataset is split into chunks and where multiple individual linear models are constructed for each chunk. 

When constructing computational models, one needs to be aware of model overfitting to the training data. Models that overfit become too tailored to the training data which reduces their generality and performance on new unseen data. Several extensions to linear models exist that can solve this for linear models. Lasso regression \cite{lasso} is one such technique. It imposes a regularization term on the number of coefficients to avoid overfitting. In particular, it uses L1 regularization by adding the absolute value of the magnitude of the coefficient as a penalty term which, when learned, shrinks the coefficients of the less important features to zero. On the contrary, ridge regression \cite{ridge_regression} applies an L2 regularization penalty by adding the square of the magnitude of the weights. 

Partial Least Squares (PLS) \cite{pls} regression is another linear method for regression problems which is especially useful when the number of predictors is higher than the number of observations, a common situation in cheminformatics, and when there is multicollinearity between the input variables. PLS tries to reduce the number of input variables to a subset that maximally is able to explain the correlation with the observed values. 

\subsection{Kernel-based Models}

One obvious drawback of linear models is their inability to fit non-linear functions to the data, which can reduce their performance on data that consists of complex, non-linear relationships between its input features and the to-be-predicted molecular properties. Several models exist that employ kernel tricks that transform the input data in such a way that it can improve the fit of linear models on these transformed data points. Support vector machines (SVMs) \cite{svm} do this by transforming the input data into higher dimensions where they become linearly separable. This transformation is achieved through kernel functions like radial basis functions. Originally SVMs were defined for classification problems \cite{svm} but were later expanded to regression problems as well \cite{svr}. Kernel ridge regression \cite{krr}, applies a similar kernel trick as in SVM to the input data but uses ridge regression to construct the linear model. While both seem to be very similar, the difference lies in the construction of the linear model. In the case of ridge regression, the construction is done by fitting the data to the linear function while in SVM it is based on the use of support vectors, which are minimal points in the data that allow describing the optimal separator function of the data \cite{svm_vs_krr}. Another simple way to fit data to output values is by using radial basis functions \cite{rbf} to approximate the unknown function. 

\subsection{Bayesian Models}
Another type of method that allows modeling non-linear relationships and does not involve more complex neural network models are Gaussian processes \cite{gaussian_processes}. These models apply principles from Bayesian inference and assume a prior probability distribution of the values of the unknown function that the Gaussian process wants to model. This distribution is updated to fit the obtained likelihood to the posterior distribution. Different from linear and kernel-based models, where a set of random parameters are fit to a fixed function form, Gaussian processes allow modeling the prior probability distribution of the model function over both all possible functional forms and their parameters. A case of Bayesian models where the functional form is fixed is the maximum likelihood estimation which uses Bayesian inference to estimate only the parameters of a chosen fixed-form probability distribution.

\subsection{Tree-based Models}

Decision tree models \cite{dt1,dt2,dt3,dt4,dt5} fit a decision tree to the training data which also consists of a feature vector of fixed size. The model uses hereby the features to construct tree nodes and tries to fit decision rules to build up the branches of the decision tree. While a single decision tree can fit all the training data, the performance might be improved when using an ensemble of decision trees each fitted to only a subset of the training data. This is what random forest models \cite{rf} do. The final prediction they generate is constructed from a weighted average of the single predictions of each decision tree in the forest. XGBoost, boosted trees and gradient boosted tree models \cite{xgboost1,xgboost2} further improve the technique of random forest models by fitting each tree sequentially instead of in parallel like in random forests and using information from the existing trees to improve the performance of the following ones. Because the performance of random forests depends on the strength of each individual decision tree \cite{rf}, this way of constructing the trees often gives improved performance over classical random forests. 

\subsection{Neural Networks}

Finally, the last group of ML models used for molecular properties prediction are neural network models. These models tend to be larger and more complex than the previously presented models and are also highly non-linear. The basic building blocks of neural network models are neurons that can be organized and connected in different ways, which gives a large variety of different types of neural network models such as feed-forward neural networks and graph neural networks.

\subsubsection{Feed-forward Neural Networks}

One of the simplest neural network models that can be used for molecular property predictions is feed-forward neural networks, such as a multilayer perceptron (MLP) \cite{mlp}. They consist of neurons which are the building blocks of neural networks and which transform the information according to the general formula \[y = \sigma(x*W + b)\] with \(x\) and \(y\) being the input and output of each neuron respectively, \(W\) and \(b\) the weight and bias assigned to each connection between neurons and \(\sigma\) an activation function applied to the neuron's output which can be either a linear or non-linear function. These neurons are further organized in layers. The input layer consists of an equal number of neurons as the number of features in the input data vector and is used to receive the input data. This is followed by one or multiple hidden layers which can be larger or smaller in size but which usually are densely connected, meaning that each neuron in a layer is connected to both all neurons of the previous and next layers. The final layers usually consist of a single output neuron that gives the scalar prediction in case of regression problems or a probability in case of binary classification. It can also consist of an output layer with multiple neurons in case of multi-class classification which gives per-class probabilities. Deep neural networks are feed-forward neural networks with a large number of hidden layers. Just as in linear models, they can contain additional tricks to improve their performance and reduce overfitting such as skip-connections, where unperturbed inputs are propagated together with their transformed counterparts to mitigate the risk of vanishing gradients, batch normalization to avoid large differences in the weights of each layer or dropout layers to avoid overfitting where a random selection of neurons are not used in specific layers.

\subsubsection{Graph Neural Networks}

So far, all of the described ML models were using 2D feature vectors to represent the molecular structural data. Graph neural networks (GNN) are a group of neural network models that can use either 2D or 3D molecular representations depending on how the input structures are presented, allowing them to explicitly use structural bond information between the atoms and 3D conformational information in the case of 3D representations. In this group of graph-based models, we can distinguish several types such as graph convolutional networks (GCN) \cite{gcn}, graph attention networks (GAT) \cite{gat}, or message passing neural networks (MPNN) \cite{mpnn}. All of these architectures model the molecular data as a 2D or 3D graph made of nodes and edges that represent atoms and bonds or interatomic distances respectively. Node and edge information is hereby represented as feature vectors of fixed size. The main difference between the different graph-based models is how they combine the information of nodes and edges in the molecular graph.

\textbf{Graph convolutional neural networks.}
Graph convolutional neural networks do this through the following propagation rule: \[H^{(l+1)} = \sigma(\tilde{D}^{-1/2}\tilde{A}\tilde{D}^{-1/2}H^{(l)}W^{(l)})\] with \(\tilde{A} = A + I_N\) being the adjacency matrix of the graph with added self connections. The adjacency matrix is a square 2D matrix with rows and columns equal to the number of nodes in the graph and it describes the connectivity between the nodes. \(\tilde{D}_{ii} = \sum_j\tilde{A}_{ij}\), \(W^{(l)}\) is the weight matrix for each layer and \(H^{(l)}\) is the matrix of neuron activations. \(\sigma(\cdot)\) is the activation function. Hereby node embeddings are updated with information from other nodes in the graph taking their graph connectivity into account. This allows the network to learn non-local, non-linear relationships by stacking multiple graph convolutional layers. 

\textbf{Graph attention networks.}
In graph attention networks, the node update operation involves an attention mechanism different from the GCNs. Concretely, the propagation rule is: \[\overrightarrow{h}_i^{'} = \Vert_{k=1}^K \sigma(\sum_{j\epsilon\mathcal{N}_i}\alpha_{ij}W\overrightarrow{h}_j)\] where \(\overrightarrow{h}_{i}^{'}\) are the updated node embeddings, \(\sigma\) is the applied non-linearity, \(\alpha_{ij}\) are the learnable attention weights, \(W\) are the weights of the neurons, \(\overrightarrow{h}_j\) the node embeddings of the current and other nodes in the graph and \(\Vert\) a concatenation across multiple attention blocks which showed better performance than when using single attention. These attention mechanisms allow the model to better learn important node connections and relations within the molecular graph and can further be manually customized by, for example, incorporating masks to only focus on local neighborhood information.  

\textbf{Message-passing neural networks.}
Lastly, message-passing neural networks instead update their node information based on neighboring nodes through message functions. Generally, message functions can be formulated as \[m_v^{t+1} = \sum_{w\epsilon\mathcal{N}(v)} M_t(h_v^t,h_w^t,e_{vw})\] with \(m_v^{t+1}\) being the total message obtained as a sum of messages coming from all neighboring nodes, \(M_t\) the individual message function operating between node embeddings of the central node \(h_v^t\) and each of its neighbors \(h_w^t\), where their connection is specified by \(e_{vw}\). These message functions are layers with learnable parameters and non-linearity and can be further customized with masks or different ways of single message aggregation. The node embeddings are then updated according to \[h_v^{t+1} = U_t(h_v^t,m_v^{t+1})\] where \(U_t\) is an update function with learnable weights and non-linearity that combines the node embeddings of each node with the message generated from their local neighborhood. These models use only local nodes that are immediately connected or that are within a predefined cutoff from the central node.

\section{Datasets}\label{sec:datasets}

In ML model training, the foundation of success lies in the utilization of extensive and top-tier training data for each specific molecular property. These invaluable datasets can be sourced either from internal, proprietary repositories or harnessed from the vast expanse of publicly available resources. In this section, we offer a comprehensive overview of curated public datasets that cater to various essential molecular properties.

\begin{table}[H]
\resizebox{\textwidth}{!}{%
\begin{tabular}{l|llll}
\textbf{Dataset name} &
  \textbf{Type of molecule} &
  \textbf{Source of molecular structure} &
  \textbf{Source of measurement} \\ \hline
\multicolumn{4}{c}{\textbf{Mixed Datasets}} \\ \hline
\href{https://go.drugbank.com/}{DrugBank} &
  Small molecules &
  SMILES &
  Experimental and predicted \\
  \\
\href{https://www.ebi.ac.uk/chembl/}{CHEMBL} &
  Small molecules &
  \begin{tabular}[c]{@{}l@{}}SMILES, target identifier, \\ PDB ID of crystal\end{tabular} &
  Experimental and predicted \\
  \\
\href{https://pubchem.ncbi.nlm.nih.gov/}{PubChem} &
  Small molecules &
  SMILES &
  Experimental and predicted \\
  \\
\href{https://tdcommons.ai/}{\begin{tabular}[c]{@{}l@{}}Therapeutic Data \\ Commons\end{tabular}} &
  \begin{tabular}[c]{@{}l@{}}Small molecules, \\ complexes and targets\end{tabular} &
  \begin{tabular}[c]{@{}l@{}}SMILES, 3D structures, \\ crystals, docking or MD\end{tabular} &
  Experimental \& assumed \\ \hline
\multicolumn{4}{c}{\textbf{Binding Affinity Datasets}} \\ \hline
\href{http://www.pdbbind.org.cn/}{PDBBind} &
  Complexes &
  Crystals &
  Experimental \\
  \\
\href{https://www.bindingdb.org/rwd/bind/index.jsp}{BindingDB} &
  Complexes &
  Crystals \& docking &
  Experimental \\
  \\
\href{http://www.bindingmoad.org/}{Binding MOAD} &
  Complexes &
  Crystals &
  Experimental \\
  \\
PLAS-5k \cite{plas5k} &
  Complexes &
  Crystals \& MD &
  Experimental \\
  \\
MISATO \cite{misato} &
  Complexes &
  Crystals \& MD &
  Experimental \\
  \\
\href{https://zenodo.org/record/5105698\#.YkLESShBxaQ}{KIBA} &
  Kinases &
  \begin{tabular}[c]{@{}l@{}}Target sequences \& \\ ligand SMILES\end{tabular} &
  Experimental \\
  \\
\href{https://dude.docking.org/}{Dud-e} &
  Targets, actives \& decoys &
  3D structures &
  Experimental \& assumed \\
  \\
\href{https://www.tu-braunschweig.de/pharmchem/forschung/baumann/muv}{MUV} &
  Targets, actives \& decoys &
  3D structures &
  Experimental \& assumed \\
  \\
\href{https://drugdesign.unistra.fr/LIT-PCBA/}{LIT-PCBA} &
  Complexes &
  Docking &
  Experimental \\ \hline
\multicolumn{4}{c}{\textbf{Physicochemical and ADMET Datasets}} \\ \hline
\href{https://www.epa.gov/tsca-screening-tools/epi-suitetm-estimation-program-interface\#citing}{PHYSPROP} &
  Small molecules &
  SMILES &
  Experimental \\
  \\
\href{http://bioinf.jku.at/research/DeepTox/tox21.html}{Tox21} &
  Small molecules &
  SMILES &
  Experimental \\
  \\
\href{https://www.epa.gov/chemical-research/exploring-toxcast-data}{ToxCast} &
  Small molecules &
  SMILES &
  Experimental \\
  \\
ClinTox \cite{clintox} &
  Small molecules &
  SMILES &
  Experimental \\
\end{tabular}%
}
\caption{Overview of popular datasets for molecular properties information used to train many ML applications. \\ 
Abbreviations: MD=molecular dynamics}
\label{tab:datasets_overview}
\end{table}

\subsection{Mixed Datasets}

Several, large-scale datasets, such as DrugBank, CHEMBL, PubChem, and Therapeutics Data Commons (TDC) exist that host data on multiple molecular properties important for drug discovery. 

DrugBank \cite{drugbank} is a dataset focusing specifically on commercially available registered drugs and their targets. The current version contains $15790$ drugs of which the majority are small molecules and $3392$ biologics. They further provide also information on the drug's commercial availability status, its mode of action, and physicochemical and ADMET properties. 
CHEMBL \cite{chembl} is a major dataset of bioactive molecular data with drug-like properties with information on various molecular properties and assays such as binding assays, ADME endpoints, toxicology information, and physicochemical properties like pKa, solubility, and lipophilicity. The dataset contains in total around $2.4$ million compounds and various amounts of assay data: $478978$ binding information datapoints, $280586$ ADME datapoints, $50784$ toxicity datapoints and $24290$ physicochemical assay datapoints. Differently from the binding datasets, CHEMBL contains the SMILES of the ligands and an identifier of the protein target or a reference to the protein-ligand complex on PDB. 
PubChem \cite{pubchem} is another major dataset of around 115 million compounds with 305 million recorded bioactivity and toxicity information. 
Therapeutics Data Commons (TDC) \cite{tdc_1,tdc_2} is an initiative and platform developed to facilitate the creation of new ML tools in various therapeutic areas. To enable this, the TDC holds in total $15 919 332$ datapoints across $66$ various datasets curated and prepared for ML model construction across 22 different prediction tasks. Apart from that, they provide additional data split functions, molecular generation algorithms, and data processing tools and hold various leaderboards to compare publicly available models and techniques on standardized benchmarks.

\subsection{Binding Affinity Datasets}\label{sec:binding_datasets}

The various datasets hosting binding affinity data can generally be divided into two groups, datasets that hold continuous binding affinity values for each protein-ligand complex and datasets based on a binary classification between binding and non-binding ligands to their targets. Datasets of the first group are PDBBind, BindingDB, Binding MOAD, KIBA, PLAS-5k and MISATO. In the second group, we have the Dud-e, Maximum Unbiased Validation (MUV) and LIT-PCBA datasets. 

\subsubsection{Datasets for Binding Affinities}

PDBBind \cite{pdbbind}, BindingDB \cite{bindingdb} and binding MOAD \cite{bindingmoad} were the first datasets containing protein-ligand complexes and experimental binding affinity values. All three have a certain degree of overlap and operate on curated subsets of the Protein Data Bank (PDB) database. 

The latest version of PDBBind currently holds 23496 complexes out of which 19443 protein-ligand complexes. This set is again divided into two parts, the general and refined sets. The refined set is a higher quality subset that was curated using a range of filters: (1) Inclusion of compounds with a resolution \(<= 2.5\mathring{A}\) and an R-factor \(<= 0.250\), (2) Exclusion of ligands with covalent bonds to the target, (3) Exclusion of complexes with multiple ligands bound in the same active site, (4) Exclusion of complexes with steric clashes (\(< 2.0\mathring{A}\)) between ligand and protein, (5) Exclusion of complexes where the ratio of the buried solvent-accessible surface of the ligand exceeds \(15\%\), (6) Exclusion of complexes with non-standard residues that are in direct contact with the ligand or complexes that have missing fragments on the backbone or sidechain of pocket residues, (7) Exclusion of complexes with ligands containing B, Be, Si and metal elements, (8) Exclusion of complexes where the ligand structure is incomplete, (9) Exclusion of complexes with large ligands exceeding a molecular weight of 1000 or contain 10 or more residues in case of peptides or peptide mimetics, (10) Only affinity data measures as constant of dissociation (Kd) or constant of inhibition (Ki), (11) Exclusion of complexes without precise binding data, (12) Exclusion of affinity data falling outside of the range 2.00-12.00 pKd/pKi, (13) Exclusion of complexes where the protein and/or the ligand in the crystal structure does not match the protein used in the binding assays, (14) Exclusion of complexes where the protein has two or more binding sites and where the bound ligands show more than 10-fold affinity differences. 

The BindingDB dataset is a larger collection of binding data holding around 2.7 million binding data points from 1.2 million compounds and 9000 targets. A vast amount of these comes from PDB crystallographic data but they also have docked target series where a set of compounds are docked to the same target with provided experimental binding affinity information. 

Binding MOAD contains crystallographic-only poses coming from PDB. It holds 41 409 protein-ligand complexes coming from $20387$ different ligands and 11058 target families. It is thus a more heterogenous dataset than the PDBBind dataset. However, from the available protein-ligand complexes, only $15223$ complexes contain binding data. 

A drawback of many binding affinity datasets and consequently ML models is that they represent the protein-ligand binding event statically through only one binding pose whereas binding has both enthalpic and entropic contributions. To overcome this limitation some groups tried to extend datasets like PDBBind with dynamic information. For this, they would run molecular dynamics (MD) simulations for the crystallographic poses in the PDBBind datasets to generate multiple binding poses and simulate the degree of movement of the ligand inside the binding pocket. This way, enthalpic contributions can be more accurately estimated and additional information can be obtained on the entropic contributions of binding. 
PLAS-5k \cite{plas5k} is one such dataset. Here they selected and simulated 5000 protein-ligand complexes from PDB and calculated several energy components from the MD data such as electrostatic, van der Waals, polar, and non-polar solvation energies.  MISATO dataset \cite{misato}  contains 20000 highly curated protein-ligand complexes from PDBBind. They used semi-empirical quantum mechanics to refine the protonation states of the complexes and fix inconsistencies in the data such as wrong element assignments. They further also computed trace information from MD trajectories as additional information on the degree of flexibility in the binding. This trace information is represented as the degree of movement of each atom across the MD trajectories. 

Besides the 3D structural data, other datasets exist comprising a collection of protein-ligand interactions with their respective affinity data without 3D binding poses. While this excludes the use of valuable 3D binding interaction information, it can provide a larger collection of affinity data and focus more on the use of other relationship information for binding affinity prediction, such as the multi-target activity of compounds. The KIBA \cite{kiba} dataset focuses specifically on kinases comprising 52498 inhibitors against 467 kinase targets. The data was merged from several studies and mapped to CHEMBL and STITCH to enable their comparison and collection of multiple binding affinity information for the same ligand-target interactions. While a direct comparison between IC50 and Kd/Ki scores cannot be performed and conversion between them depends on substrate concentration information (Cheng-Prusoff model \cite{cheng_prusoff}: \(K_i=IC_{50}/(1+[S]/K_m)\)) that often is lacking in reported binding affinity data, the authors noted on the existence of correlations between \(IC_{50}\) and \(K_d\)/\(K_i\) data in CHEMBL. They used therefore adjustments to the reported \(K_d\) and \(K_i\) scores based on reported \(IC_{50}\) values for the respective protein-ligand interactions. These adjusted \(K_d\) and \(K_i\) scores were also merged in case both were reported for the same protein-ligand interaction. These adjusted and combined affinity scores, called KIBA scores, span a range describing both binding and non-binding interactions. 

\subsubsection{Binding Datasets for Classification}

All of the previously mentioned binding affinity datasets with the exception of the KIBA dataset provide positive binding information. While this is certainly very valuable information, ML models generally benefit from rich and heterogenous data to learn complex patterns. Therefore, a possible issue of ML models trained on such binding data is their inability to detect non-binding ligands for which one can obtain reasonably good docking poses \cite{decoys_good_docking}. This is important since, in practice, virtual screening datasets are comprised of a mixture of potential binders and non-binders. Thus, many ML models for binding affinity are unusable despite showing high performance on binder compounds. In addition, real virtual screening datasets show also a strong unequal distribution in favor of non-binders, which makes the prediction task and selection of the top binding compounds harder. To enable the detection of possible non-binding, decoys, molecules specific datasets have been constructed comprising of both positive binders as well as decoys, such as the Dud-e \cite{dude}, MUV \cite{muv} and LIT-PCBA \cite{lit_pcba} datasets. 

The Dud-e dataset \cite{dude} is one of the widely used datasets for binding classification. The dataset is a revised and improved version of the preceding Dud dataset \cite{dud} that had several internal biases \cite{dud_bias_1,dud_bias_2,dud_bias_3}. It comprises both 22886 active compounds with activities against 102 targets and 50 generated decoys per active compound. The decoys are generated in such a way that they have similar physicochemical properties as the actives but different 2D topology. Some \cite{bias_dude_1,ave,bias_dude_3,bias_dude_4}, however, have addressed that the Dud-e dataset still has biases between the active and decoy compounds such as the difference in 2D topology, which can be an easy discriminator for ML models to capture. This can lower the practical usability of the trained models. Therefore, better algorithms and datasets have been proposed to overcome these biases, such as the Maximum Unbiased Validation (MUV) or the LIT-PCBA dataset. 

The MUV dataset \cite{muv} was specifically curated from data taken from PubChem and consists of 15 target subdatasets with 30 active and 15000 decoy molecules each. The data curation consisted of selecting active and inactive compounds confirmed experimentally through both primary and confirmatory bio-activity screens. The actives were further filtered for unwanted compounds such as frequent hitters, high aggregations, or compounds with chromo/fluorogenic properties for screening assays based on optical detection methods. Quality checks on the decoys most similar to the actives were also performed by checking literature sources for any potential binding to the respective target in order to exclude potential false negatives. Lastly, to overcome the structural biases between active and decoy compounds seen in datasets like the Dud-e dataset, the creators of the MUV dataset employed chemical space embedding filters to remove both actives not properly embedded in the decoy chemical subset space and vice versa. This ensures that apart from physicochemical properties the actives and decoys are also structurally similar. Despite the extra effort to reduce bias, it was pointed out \cite{ave} that also this dataset has internal bias with decoys not having a proper homogeneous distribution in the chemical space making decoys easy to classify and detect. 

Another dataset that was constructed to adjust for the different biases found in the previous datasets is the LIT-PCBA dataset \cite{lit_pcba}. The dataset contains in total of 15 targets with 7844 actives and 407381 inactives. They applied a similar approach as the authors from the MUV dataset, using data from PubChem BioAssays to select confirmed actives and non-active compounds and applying a series of filters to remove non-drug like compounds, compounds with undesired physicochemical properties and compounds that are known to give false positives in many assays such as frequent hitters, compounds with chromo/fluorogenic properties and compounds giving high aggregations in assays. They further selected compounds most similar to compounds found in the PDB database and generated several conformers for each selected compound. The most similar conformer to the PDB ligand was selected for each target set. All compounds were further docked to their respective targets. To overcome the biases found in the previous datasets they used the asymmetric validation method (AVE) \cite{ave} to ensure an unbiased selection of actives and inactives in each target sub-dataset. This method measures the pairwise similarities of compounds that belong to one of the 4 subsets (training actives, validation actives, training inactives, validation inactives) and attempts to select training and validation compounds that give the lowest bias scores. 

\subsection{Physicochemical and ADMET Datasets}

Various datasets exist related to physicochemical molecular properties and ADMET, like PHYSPROP, Tox21, ToxCast, and ClinTox. 
The PHYSPROP dataset \cite{physprop} contains information on 13 physicochemical and environmental fate properties including octanol/water partition coefficients. In total, it contains 47047 chemicals out of which 15806 have octanol/water partition coefficient information. 
Tox21 \cite{tox21_1,tox21_2} is a dataset comprising $12707$ data points as chemical compounds and results on 12 toxicological endpoints. For each compound in the dataset, $801$ dense and $272776$ sparse features are included that represent chemical descriptors and chemical substructures respectively. The dataset also comes with training and test subsets making it readily available for machine learning. 
Another dataset for toxicity information is the ToxCast dataset \cite{toxcast}. This dataset contains around 8000 compounds with results about toxicity on over 600 endpoints. All the data is presented in binary format representing the existence of toxicity against a specific endpoint marker. 
ClinTox \cite{clintox} is an interesting dataset that contains toxicity information from successful and failed clinical trials for $1484$ drugs. All the negative data was collected from the database for Aggregate Analysis of Clinical Trials (AACT) at ClinicalTrials.gov where only drugs were selected that failed the clinical trial for toxicity reasons. The positive data was selected from DrugBank as FDA-approved drugs.

\section{Molecular Properties}
\subsection{General Overview}

\begin{figure}[H]
\centering
     \includegraphics[width=1.0\textwidth]{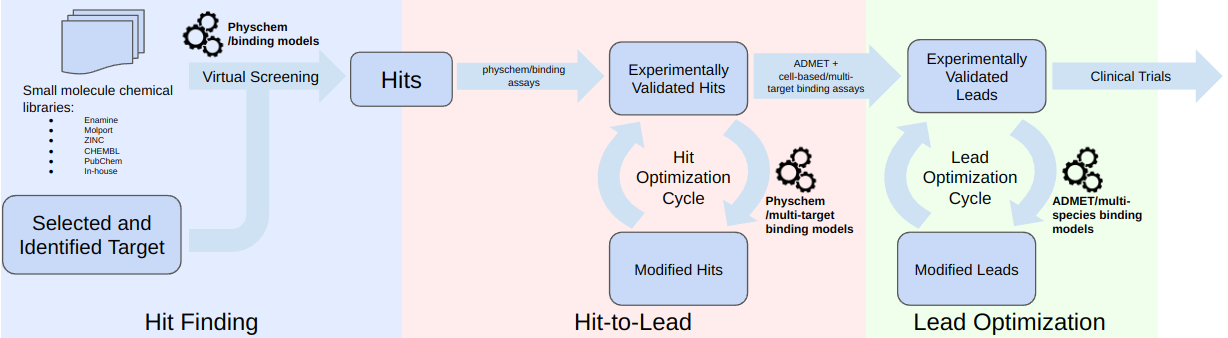}
      \caption{Workflow of computational hit finding and lead generation. During hit finding, large libraries of small molecule purchasable compounds are screened through predictive models for physicochemical and binding affinity properties against the specific target of interest. These best-scoring hits are further validated through physicochemical and binding assays. Once validated hits are obtained with moderately desired molecular properties, these can further undergo optimization through multiple consecutive rounds of modifications, followed up by virtual screening with predictive models and experimental validation. Important properties to optimize hereby are their physicochemical properties and binding affinity against the main target while ensuring low binding affinity against similar targets to ensure the compound's selectivity. Once optimized hits are obtained for these molecular properties, they undergo more thorough experimental validation using more expensive cell-based and multi-target binding assays and ADMET property assays. The obtained best-scoring lead candidates further undergo a second optimization cycle to improve their ADMET-related properties and ensure target selectivity and second-species validation.}
       \label{fig:lead_schema}
\end{figure}

When looking at the classical pipeline of small molecule drug discovery and development in Figure \ref{fig:lead_schema}, we can see at which stages different machine learning based scoring functions can be applied for different properties that need to be established. Additionally, such classical pipelines can also become completely semi-automatic \cite{drugsniffer}. For this, a combination needs to be made of different computational techniques such as ML models for molecular property predictions, docking software and target and binding pocket identification methods. 

Models predicting various physicochemical properties of the selected or generated small molecules are among the first models that are applied. This is to ensure that the molecules selected for binding affinity analysis have already the desired physicochemical properties such as correct protonation state and solubility. 

Alternatively one could also screen for binding affinity together with the prediction of the physicochemical properties when using simple models that take only separate protein and ligand information into account (see Section \ref{sec:sep_protlig}). However, when using models that operate on 3D bound protein-ligand complexes for binding affinity prediction (see Section \ref{sec:protlig_complex}), it is advisable to employ physicochemical property predictors beforehand to reduce the number of molecules that would go into binding affinity prediction. This is because these models require the molecules to be bound to their target, making docking a bottleneck in the pipeline as ML models in general are capable to produce results instantaneously. 

Finally, ADMET models can be applied at a later stage to ensure that optimized leads have favorable absorption, distribution, metabolism, excretion, and toxicity profiles. In principle, these models could also be applied during the initial stages together with the physicochemical predictive models, however, this could potentially reduce the chemical space of compounds tested for binding affinity. One needs to remember that in a pipeline as presented in Figure \ref{fig:lead_schema}, experimental testing is performed each time after binding affinity and ADMET predictions of the studied compounds and selection of the top-scoring ones, to experimentally validate the selected compounds. Therefore, using ADMET predictive models early in the pipeline could first of all reduce the diversity of compounds selected for binding affinity estimation leading to a sub-optimal exploration of the chemical space of compounds that can potentially bind well to the target. Second, it would require additional experiments on ADMET profile estimation early on a larger selection of molecules. Taking the higher cost of these experiments into account, this could result in being cost-ineffective. Lastly, one can also observe that two cycles exist in the pipeline for hit and lead optimization. These cycles represent consecutive compound optimization, their screening with predictive models, and experimental validation in order to further improve and select compounds with better molecular properties.

\subsection{Physicochemical Properties Predictive Models}\label{sec:physchem_props}

Physicochemical properties such as solubility, lipophilicity, and pKa \cite{physchem_props_DD} are important properties that need to be established early in the hit and lead generation cycle. As they can influence binding affinity and ADMET properties, it is advisable to establish them early to minimize drug attrition in the subsequent stages. Traditionally these properties have been measured through classical wet-lab experiments such as spectrographic analysis or capillary electrophoretic mobility for pKa determination, plate partitioning, reversed-phase high performance liquid chromatography (HPLC) or capillary electrophoresis for lipophilicity and UV or light scattering for solubility measurements \cite{physchem_props_DD}. Various computational techniques have also been developed to predict these properties from empirical algorithms to more advanced machine learning models. Some advancements in the latter are shown in Table \ref{tab:physchem_overview} and will be discussed further in this section.

\begin{table}[H]
\centering
\resizebox{\textwidth}{!}{%
\begin{tabular}{l|l}
\textbf{Embedding type}             & \textbf{Tested models}                                                        \\ \hline
\multicolumn{2}{c}{\textbf{pKa Prediction Models}}                                                                  \\ \hline
Rooted topological torsion FPs &
  RF, PLS, XGBoost \textbf{\underline{Lasso}}\cite{pka_rooted_top_fp}, SVR \\ 
  \\
Molecular descriptors &
  \textbf{\underline{RBF NN}}\cite{pka_pso_rbfnn}, SVM, XGBoost, DNN \\
  \\
Molecular + structural descriptors &
  \begin{tabular}[c]{@{}l@{}}\textbf{\underline{SVM}}\cite{pka_qsar}, XGBoost, \textbf{\underline{DNN}}\cite{pka_qsar}, \\ \textbf{\underline{RF}}\cite{ml_meets_pka}, SVR, MLP, XGBoost\end{tabular} \\
  \\
DFT descriptors & 
  \begin{tabular}[c]{@{}l@{}}PLS, \textbf{\underline{RBF NN}}\cite{pka_qm_rbf}, RF, \\ Gaussian processes, \textbf{\underline{KRR}}\cite{pka_dft}, feed-forward NN\end{tabular} \\
  \\
Molecular graph & 
  \begin{tabular}[c]{@{}l@{}} \textbf{\underline{MPNN}}\cite{mf-sup-pka,molgpka}, \\ \textbf{\underline{AttentiveFP}}\cite{multi-instance_pka}, \textbf{\underline{GCN}}\cite{epik}\end{tabular}                    \\ 
  \hline
\multicolumn{2}{c}{\textbf{Aquaeous Solubility Prediction Models}}                                                                    \\ \hline
Molecular + structural descriptors  & \textbf{\underline{XGBoost}}\cite{aq_sol_adme_knime} \\
\\
Molecular graph                     & \textbf{\underline{GCN}}\cite{admet_aq_sol}, \textbf{\underline{GAT}}\cite{lipo_admet_gnn}, MPNN, AttentiveFP         \\ \hline
\multicolumn{2}{c}{\textbf{Lipophilicity Prediction Models}}                                                                          \\ \hline
Molecular descriptors               & \textbf{\underline{RF}}\cite{generative_mmpa_qsar_admet}, SVM, XGBoost, \textbf{\underline{MLP}}\cite{lipo_rt}  \\
\\
Molecular + structural descriptors &
  RF, XGBoost, MLR, \textbf{\underline{FFNN}}\cite{multitask_dnn_admet} \\
  \\
Molecular graph & 
  GCN, \textbf{\underline{GAT}}\cite{lipo_admet_gnn}, MPNN, AttentiveFP        
\end{tabular}%
}
\caption{Overview of predictive models for pKa, solubility and lipophilicity prediction. Underlined the best-performing models from each reviewed publications. \\
Abbreviations: GNN=Graph Neural Network, RF=Random Forest, PLS=Partial Least Squares, XGBoost=Extreme gradient Boosted Trees, SVM=Support Vector Machine, KRR=Kernel Ridge Regression, NN=Neural Network, RBF=Radial Basis Function, kNN=k-Nearest Neighbours, DNN=Deep Neural Network, GCN=Graph Convolutional Neural Network, GAT=Graph Attention Network, MPNN=Message Passing Neural Network, FP=Fingerprint, MLR=Multivariate Linear Regression, FFNN=feed forward neural network}
\label{tab:physchem_overview}
\end{table}

\subsubsection{pKa Predictive Models}\label{sec:pka}

It is important during early screening and novel compound generation, to assign the molecules their correct protonation state at physiological pH. For this, detection of the protonation sites and accurate pKa prediction for each is important. This should happen prior to the prediction of any other property due to the influence of the molecule's protonation state \cite{prot_state_importance} on any other prediction. Different models have been developed using various ways to embed the chemical information presented to the ML models (Table \ref{tab:physchem_overview}). Common ways to embed the molecules are through chemical and structural descriptors or molecular network graphs. 

\textbf{Chemical and structural descriptors.}
Molecular chemical and structural descriptors are a type of embedding that constructs a 2D feature vector encoding physicochemical or structural properties of the compounds through different techniques such as rooted topological fingerprints, molecular property embeddings, structural fingerprints or QM-based descriptors. 

Lu et al. (2019) \cite{pka_rooted_top_fp} uses rooted topological fingerprints to generate embeddings that can be coupled with various ML models. These fingerprints are specifically designed to capture structural information around a central atom. They can do this in 3 ways: (1) through RPairs \cite{atom_pair} which embed atom pairs starting from the central root atom and that can be located at n-bonds distance from each other, (2) through RTorsions \cite{rtorsion} which is a path-based fingerprint method that constructs the embedding vector from paths starting at the central root atom and a maximum n-bonds distance, (3) RMorgan \cite{pka_rooted_top_fp} which is based on the Morgan fingerprint \cite{morgan_fp}, a type of circular fingerprint that embeds information from the central root atom and all the neighboring atoms within a radius of n-bonds. The n distance is each time specified by the user. In all the 3 ways the atoms are defined by a short vector that holds different types of information. In RPairs this vector embeds information about the atomic element, the number of bonded non-hydrogen atoms, and the number of bonding \(\pi\) electrons that it has. In RTorsion the vector embeds information about the atomic element, the number of \(\pi\) electron pairs between the consecutive atoms in the path and the number of non-hydrogen atom neighbors that are not in this path. In RMorgan the vector embeds information about the atomic element, the number of non-hydrogen atom neighbors, the number of attached hydrogen atoms, the atomic charge, isotope information and ring membership. 

In Lu et al. (2019) \cite{pka_rooted_top_fp} they combined the constructed fingerprints with random forest, XGBoost, PLS, Lasso regression, SVM or kNN models. They saw that the RTorsion embedding method provided the best results coupled with Lasso regression. XGBoost and SVM showed also close to top performance on their time-split-based test set. The embedding described here is a structural descriptor type embedding as it tries to use information on the structural arrangement of the molecule. 

Another group of descriptors that can be used are chemical descriptors which provide a series of calculated molecular properties such as molecular weight, logP/logD, quantitative estimate of drug-likeness (QED) \cite{qed}, number of hydrogen bond donors or acceptors and many others. All these can be easily computed using the RDKit \cite{rdkit} Python library. In Baltruschat et al. (2020) \cite{ml_meets_pka} they used 196 molecular descriptors together with a 4096-bit long structural Morgan fingerprint \cite{morgan_fp} with radius 3 as embedding vector coupled with random forests, SVMs, MLP and XGBoost models. Morgan fingerprints coupled with random forests showed the best performance on two test sets, the publicly available Novartis test set \cite{novartis_pka_test} and the own curated set from several literature sources \cite{avlilumove_pka_test1,avlilumove_pka_test2,avlilumove_pka_test3,avlilumove_pka_test4,avlilumove_pka_test5}. While they report good accuracies on the literature test set, this method has the disadvantage that pKa values for each specific protonation site of the molecule cannot be predicted, as the whole molecule is embedded into a single vector. Therefore this method cannot be used for structures bearing multiple protonation sites like amphoteric molecules. 

Mansouri et al. (2019) \cite{pka_qsar} use also a combination of molecular descriptors, binary structural fingerprints, and fragment counts. The latter splits the molecule into smaller fragments and generates a fingerprint based on the number and type of fragments in the whole molecular structure. They coupled this embedding with SVMs and kNN models. While their approach also does not make it possible to generate predictions for individual protonation sites, they solve the problem of amphoteric molecules by applying a step-wise prediction approach using both classification and regression models. They first build classification models that classify the molecules into one of three groups: (1) acidic molecule, (2) basic molecule, and (3) amphoteric molecule. The acidic and basic groups are established in the same way as in MolGpka \cite{molgpka}. For each class, they construct separate regression models predicting the basic and acidic pKa separately. 
While such an approach allows to generate predictions for amphoteric molecules containing a basic and acidic protonation site, they still cannot be used for molecules with more complex protonation profiles, containing multiple protonation sites. As seen in Lu et al. (2019) \cite{pka_rooted_top_fp}, in order to be able to generate predictions for multi-protic compounds, it is important to generate local embeddings that take information primarily from the local neighborhood of the protonation site. Where in Lu et al. (2019) \cite{pka_rooted_top_fp} this was achieved through structure-based fingerprints, Hunt et al. (2020) \cite{pka_qm_rbf} use QM chemical descriptors to embed each protonation site of the input molecule. The used descriptors capture the atomic and bond properties of the atom bearing the protonation site with its surrounding bonded hydrogen and heavy-atom neighbors. Properties such as nucleophilic and electrophilic delocalizabilities, bond lengths and atom charges were used together with information on the highest occupied molecular orbital (HOMO) and lowest unoccupied molecular orbital (LUMO) energies and heats of formation. For polyprotic compounds, most acidic and basic sites were first established following a step-wise approach to locate other protonation sites while maintaining the correct protonation for already established sites. This allows for a more accurate calculation of the QM descriptors. They further tested the generated embedding with linear, partial least squares (PLS), radial basis functions (RBF), random forest (RF) and Gaussian process (GP) ML models where RBF, GP and RF models showed the best performance. 

In Lawler et al. (2021) \cite{pka_dft} both protonation-centric and whole molecule chemical descriptors are used to construct the molecular embedding such as information on the electronegativity of the central atom, magnitude of the dipole moment of the whole molecule, the degree of oxidation of the central atom, number of hydrogens in the whole molecule, number of fluorine and carbon atoms which say something about major electron-withdrawing groups and size of the overall molecule respectively, the molecule's molecular weight, the Connolly volume \cite{connolly_volume} of the molecule, solvation free energy and DFT calculated pKa values for the specific protonation site. By using a DFT-computed pKa value as a molecular descriptor, this method could be seen as a further refinement of the DFT-calculated pKa values by taking other types of chemical information into account. These molecular embeddings are further coupled with ML models such as kernel ridge regression (KRR), Gaussian processes and feed-forward neural networks. Hereby KRR gave the best performance and reported a lower error with the experimental pKa values than the initial DFT-calculated pKa values.

\textbf{Molecular network graph embedding.}
Graph-based models use node and edge chemical feature embeddings with a GNN. The input is a small molecule that is transformed into a network graph with nodes and edges as atoms and bonds respectively. Each node and edge is further embedded through 2D vectors that can be constructed according to different rules. 

MolGpka \cite{molgpka} uses 1-hot embedding for the nodes consisting of a total of 39 bits that encode information such as the atom element, its hybridization state, whether the atom is a hydrogen bond donor or acceptor, the atom's degree, its valence, whether it forms part of a cyclic structure and the size of the ring, whether the atom is aromatic and whether it is the protonation site. For the edges, they use a binary adjacency matrix representing the connectivity of the nodes according to existing chemical bonds. 

Graph-pKa \cite{multi-instance_pka} uses also 1-hot encoded node embeddings similar to MolGpka \cite{molgpka}. The difference between their node embeddings lies in a different number of bits for atom element information, the atom's degree and the hybridization state of the atom. Further MolGpka \cite{molgpka} encodes information about hydrogen donating and accepting properties of the atom and whether the atom is the protonation site which is not used in Graph-pKa \cite{multi-instance_pka} node embedding. While the latter embeds information on the formal charge, presence of radical electrons, number of bonded hydrogens and chirality. Different from MolGpka \cite{molgpka} they also apply a 1-hot encoded embedding vector on the edges which are taken to be chemical bonds between the atoms. The edge embedding vector encodes information about the type of bond, whether it is conjugated or not, whether it is part of a cyclic structure and stereo-chemical information. 

Similar to this, MF-SuP-pKa \cite{mf-sup-pka} uses a 40-bit long node embedding vector based on Xiong et al. (2020) \cite{multi-instance_pka_model} adding 1 additional bit for the atomic degree information. They also apply a 1-hot encoded embedding vector for the edges based again on the one used in Xiong et al. (2020) \cite{multi-instance_pka_model} and add 2 additional bits for the stereo-chemistry information. 

Epik \cite{epik} also uses a node embedding vector based on chemical information. They use a 74-bit long vector encoding information about the atomic element, its degree, the number of valence and radical electrons, the formal charge, the hybridization state of the atom, whether it is aromatic and the number of explicit bonded hydrogens. The main difference between their embedding vector and the previous ones is that they encoded a larger number of possible atomic elements, use more bits to represent the atom's degree and use a mixture of 1-hot encoded and single continuous values in the embedding. Similar as in MolGpka \cite{molgpka} they establish edges based on bond connectivity of the atoms. 

Once nodes and edges are embedded, information is exchanged between the nodes through either message-passing layers like in Graph-pKa \cite{multi-instance_pka} and MF-SuP-pKa \cite{mf-sup-pka} or graph convolutions like in MolGpka \cite{molgpka} and Epik \cite{epik}. Both Graph-pKa \cite{multi-instance_pka} and MF-SuP-pKa \cite{mf-sup-pka} use the attention-based message passing architecture from Xiong et al. (2020) \cite{multi-instance_pka_model}. This uses a node-level attention vector constructed from a weighted combination of the node's neighbors which is merged with the node's updated embedding vector at each layer through gated recurrent units (GRUs) which help to take influences of further away located nodes into account. Hereby, edge embedding vector information is first added to the node embedding vectors each time prior to the attention-based message passing operation. 

While Graph-pKa \cite{multi-instance_pka} uses this type of message passing on the complete molecular graph centered around the atom bearing the protonation site, MF-SuP-pKa \cite{mf-sup-pka} defines first k-hop molecular substructures around each atom that holds the protonation site and applies an intermediate weighted pooling operation within the nodes of each molecular substructure followed by additional attention-based message passing operations on the pooled super-nodes of each molecular substructure. 
Different in MolGpka \cite{molgpka} from Epik \cite{epik} is that in the prior, two separate networks are used for acidic and basic protonation sites respectively. This is done in order to keep the input molecule in its neutral state while in Epik \cite{epik} the input molecule's protonation site(s) is/are always in the protonated form with respect to their experimental pKa value. 
Finally, all models use fully connected layers to reduce the embedding vector of the node that holds the protonation center to the scalar prediction value for the pKa. 

Traditionally, neural network models are trained by backpropagating the gradient of the loss to each parameter in the network. CSAPSO-EDCD RBF ANN \cite{pka_pso_rbfnn} uses an optimized particle swarm optimization (PSO) algorithm \cite{pso1,pso2} to select the input features and train their neural network. PSO is a type of genetic algorithm that tries to optimize a set of parameters in parallel through a search for the best parameter combination taking into account the value of the other parameters. Different from the other methods, they do not model the molecules as graphs but instead, compute a series of 686 molecular descriptors that are further reduced to 5 using their improved PSO algorithm. These were then coupled with a radial basis function (RBF) neural network, which is a type of feed-forward neural network typically characterized by an input layer, 1 hidden layer that uses radial basis functions as activation functions and 1 linear output layer. In this work, PSO was further applied to find the variables of the RBF functions of each neuron that fit the data. Different from the previous methods, this method does not embed each protonation site inside the molecule but instead generates a single whole molecule embedding and would therefore fail on multi-protic compounds. 

\textbf{Performance comparison.}
In order to compare the performance of models it is important to use standard benchmark test sets to make the comparison as least biased as possible. However, benchmark test sets used in the different works on pKa prediction show different test sets used to benchmark their models. This makes direct comparisons between them hard. In general, all methods report high performance with squared Pearson's correlations above 0.90. Baltruschat et al. (2020) \cite{ml_meets_pka} does report lower performance on the Novartis test set \cite{novartis_pka_test}, probably because of the higher heterogeneity and size of the compounds in the test set. This indicates that some of the models have certain limitations as discussed above in terms of generating predictions for more complex multi-protic compounds. Therefore, methods that model each protonation site separately can be more universally applied to different small molecules as opposed to methods that generate a single embedding for the whole molecule. Interestingly, despite being more simple, non-neural network methods are able to also achieve competitive performance as more complex graph-based neural network models.

\subsubsection{Aqueous Solubility Predictive Models}

After adjusting the molecules to the correct protonation states one of the other important physicochemical properties to predict is the aqueous solubility, logS (in mol/L). As this property is generally an effect of the whole molecular structure rather than local molecular neighborhoods like in pKa prediction, one can easily get good performance by applying whole molecule chemical and structural embeddings coupled with simple linear or decision tree-based models. 

Falcòn-Cano et al. (2020a) \cite{aq_sol_adme_knime} uses $1400$ whole molecule chemical descriptors together with $45$ additional physicochemical properties. To reduce the number of features in the generated embeddings they applied a selection by permutation of the variables using a random forest model where only high occurring variables were selected in the individual decision trees of the random forest model, together with recursively selecting the most correlated variables. This was further coupled with both classifier and regression XGBoost models as solubility can span a wide range of values. To do this, training data was classified into a soluble and highly soluble class (logS \(\ge\) -2) and a slightly soluble and insoluble class (logS \(<\) -2) to be used to train the classifier model. Separate regression models were further trained for the two separate classes. Additionally, a third regression model was trained on all the training data and an ensemble approach was used by taking the average of the local and global regression models. The method was tested against two external test sets with curated data from the literature. The performance of the final regression model showed a median performance of around 0.64 and 0.69 Pearson's correlation for models trained on cleaned data points only and extended with reliable data points respectively based on the accuracy of the reported experimental logS values on the test set 1 and 0.43 for test set 2. The performance of the classifier model showed a good performance of 0.80 accuracy, 0.60 Cohen's Kappa, 0.89 sensitivity and 0.71 specificity for test set 1 and 0.83 accuracy, 0.67 Cohen's Kappa, 0.73 sensitivity and 0.93 specificity for test set 2 when using only cleaned data points. 

Different from this, Chemi-Net \cite{admet_aq_sol} uses a graph-based neural network, similarly as described in \ref{sec:pka}. They used both atom and bond chemical descriptor-based embedding vectors containing information about the atom type, van der Waals and covalent radius of the atom, the number of rings the atom belongs to, whether the atom is in an aromatic ring or not, and the electrostatic charge of the atom for atom embeddings. For bond embeddings they used the bond type, bond length, and whether the bond is part of a ring system. They used a convolutional graph neural network to update the atom and bond embeddings and applied several pooling and dimensionality-reducing layers to generate the output value. Interestingly, they trained their model on several molecular properties in parallel such as aqueous solubility, CYP450 inhibition, human liver microsomes, bioavailability, and PXR induction. All these other properties are ADMET specific and will be discussed further in Section \ref{sec:admet}. To train the model in such a parallel fashion, they applied a combined loss function over the individual predictions. This allows the model to learn across datasets and improve performance on each sub-dataset especially when few data points are available. The performance varies depending on the molecular property predicted with Pearson's correlations ranging between 0.11 and 0.692 when the model is trained on one single task only. When trained with the multi-task loss, performance improves for some of the properties, including the low-performing ones for which performance increased for example from 0.2 to 0.327 Pearson's correlation. This shows that multi-task learning can help when a few data points are available for one task. Also, performance seems to correlate only slightly with training data size. Computing Pearson's correlation between the obtained performances on the test sets for the single-task learning and the training set sizes gave a correlation of 0.136. A similar correlation was obtained using test set sizes. This indicates that amount of training data has importance to some degree but that data quality is equally if not more important. 

\subsubsection{Lipophilicity Predictive Models}\label{lipophilicity}

A third important physicochemical property that needs to be established is lipophilicity as it has important implications for the molecule's solubility and membrane passage. This is expressed as partition coefficients of the compounds between a hydrophobic and a hydrophilic environment either as logP values for non-ionizable compounds or logD values for ionizable compounds where the distribution of the compound between the two phases depends on the fraction of the ionized and non-ionized species which is influenced by the surrounding pH. Again, just as in aqueous solubility, the property depends on the whole molecule. Therefore, whole molecule chemical and/or structural descriptors can be generated and coupled with different ML methods. 

Win et al. (2023) \cite{lipo_rt} used 204 chemical descriptors from RDKit \cite{rdkit} which were further pruned to 125 excluding descriptors with low variance, high correlation to other descriptors and descriptors with missing and zero values. These chemical descriptors were further augmented with structural Morgan fingerprints \cite{morgan_fp} and experimental reversed phase chromatography retention times. These embeddings were further tested with several ML methods such as SVM, MLP, XGBoost and random forests. They constructed separate models to predict both logP and logD values. The MLP model gave the best performance on the validation set and gave high performance on the test set with Pearson's correlation values above 0.85 on both metrics. From further feature interpretability they found unsurprisingly that the retention time had the most impact on the predictions. 

Just as in Chemi-Net \cite{admet_aq_sol}, in Wenzel et al. (2019) \cite{multitask_dnn_admet} they used multi-task learning to train a deep neural network using chemical molecular descriptors with atom-pair and pharmacophoric donor-acceptor pair descriptors. They train the model on a set of different molecular properties such as metabolic clearance, passive permeability in Caco-2 cells, metabolic liability, and logD values. Different from Chemi-Net \cite{admet_aq_sol}, they trained their model in a step-wise manner by training the model on one task, optimizing the shared weights and task-specific weights of the neural network model and keeping weights for other tasks frozen. This makes it possible to train the network by using different independent datasets with a limited overlap of compounds. From the results, they show that by combining related datasets where the properties have a certain relationship such as data on metabolic liability in different animal species, performance improves compared to a single-task trained model. However, combining unrelated datasets such as metabolic liability data with logD does not always give an improvement in performance. In general, performances range from around 0.65 squared Pearson's correlation for metabolic liability datasets to around 0.85 for logD prediction with data from in-house company experimental screening showing more consistency than when using publicly available data from CHEMBL, where performances could drop to around 0.50 on some benchmarks. 

In Broccatelli et al. (2022) \cite{lipo_admet_gnn} again they trained graph-based models in a multi-task approach using data for different molecular properties such as logD, intrinsic clearance in human liver microsomes and hepatocytes, and kinetic solubility in a phosphate buffer. They apply 1-hot encoded chemical-based atom and bond embeddings similar to some previously seen graph-based models with information on the atom type and its degree, whether the atom is chiral, its formal charge, hybridization state, number of implicit valences, whether the atom is aromatic, the bond type, whether the bond is conjugated and part of a ring system and stereo configuration of the bond. They used different types of graph-based models such as graph convolutional networks, graph attention networks, message passing networks, and the attentive fingerprint model from Xiong et al. (2020) \cite{multi-instance_pka_model}. For the multi-task prediction, they tested two approaches: (1) using task-shared and task-specific layers in the neural network models such as in Chemi-Net \cite{admet_aq_sol} and (2) a bypass architecture where separate neural networks are trained for each single task and a general model trained for all tasks. The final prediction is then the ensemble of the output of both the task-specific and general models. The graph attention network showed better performance in single-task learning but multi-task learning did not show always an improvement over single-task learning. It had small improvements for prediction of molecular properties such as solubility, metabolic liability and clearance which can benefit from information such as lipophilicity, but not vice versa. In general, performances ranged from 0.30 to 0.63 squared Pearson's correlation for the different properties evaluated on time-split-based test sets indicating average performance.
 
\subsection{Binding Affinity Predictive Models}

The next important property that needs estimation is the binding affinity of a complex. The binding prediction can be performed in three different ways: (1) as a classification where the model classifies compounds as binders or non-binders or into different binding affinity ranges; (2) absolute, where the model predicts the binding affinity metric directly of the molecule to its target; (3) as a prediction of the relative binding affinity between pairs of compounds binding the same target. Further, the three ways will be discussed together with their existing models.

{\small\tabcolsep=3pt
\begin{longtable}{p{5cm}|p{6cm}p{4cm}}
\textbf{Method name} &
  \textbf{Embedding type} &
  \textbf{Tested models} \\ \hline
\endfirsthead
\multicolumn{3}{c}%
{{\bfseries Table \thetable\ continued from previous page}} \\
\textbf{Method name} &
  \textbf{Embedding type} &
  \textbf{Tested models} \\ \hline
\endhead
\multicolumn{3}{c}{\textbf{Binding Affinity Classification}} \\ \hline
Morris et al. (2020) \cite{classbin5}** & 
  \begin{tabular}[c]{@{}l@{}}Text-based transformer \\ embedding\end{tabular} & 
  \textbf{\underline{FFNN}} \\
  \\
Torng et al. (2019) \cite{classbin6}* & 
  \begin{tabular}[c]{@{}l@{}}Protein pocket \& \\ ligand graph embedding\end{tabular} & 
  \textbf{\underline{GCN}} \\
  \\
vScreenML \cite{classbin1} &
  Rosetta energy terms &
  \textbf{\underline{XGBoost}} \\
  \\
Nogueira et al. (2019) \cite{classbin2} &
  PADIF interaction FPs &
  FFNN, \textbf{\underline{SVM}} \\
  \\
BindScope \cite{classbin3} &
  3D voxels &
  \textbf{\underline{3D CNN}} \\
  \\
Lim et al. (2019) \cite{classbin4} &
  Molecular graph embedding &
  \textbf{\underline{GAT}} \\ \hline
\multicolumn{3}{c}{\textbf{Absolute Binding Affinity}} \\ \hline
ChemBoost \cite{absbin17}* &
  \begin{tabular}[c]{@{}l@{}}Ligand SMILES embedding +\\ ligand-based protein embedding\end{tabular} &
  \textbf{\underline{XGBoost}} \\
  \\
DeepFusionDTA \cite{absbin19}* &
  \begin{tabular}[c]{@{}l@{}}Ligand SMILES embedding + \\ protein sequence embedding\end{tabular} &
  \textbf{\underline{light GBM}} \\
  \\
AttentionDTA \cite{absbin33}* &
  \begin{tabular}[c]{@{}l@{}}Ligand SMILES embedding + \\ protein sequence embedding\end{tabular} &
  \begin{tabular}[c]{@{}l@{}}\textbf{\underline{1D CNN}} \\ \textbf{\underline{with attention}}\end{tabular} \\
  \\
DeepDTA \cite{absbin39}* &
  \begin{tabular}[c]{@{}l@{}}Ligand SMILES embedding + \\ protein sequence embedding\end{tabular} &
  \textbf{\underline{1D CNN}} \\
  \\
SimCNN-DTA \cite{absbin21}* &
  \begin{tabular}[c]{@{}l@{}}Ligand-ligand and protein-protein\\ similarities\end{tabular} &
  \textbf{\underline{2D CNN}} \\
  \\
ECIF-LD-GBT \cite{absbin9} &
  \begin{tabular}[c]{@{}l@{}}ECIF + ligand chemical \\ descriptors\end{tabular} &
  \textbf{\underline{XGBoost}} \\
  \\
Wang et al. (2021a) \cite{absbin13} &
  Proteo-chemometrics IFP &
  \begin{tabular}[c]{@{}l@{}}\textbf{\underline{RF}}, GBDT, FFNN,\\ DT\end{tabular} \\
  \\
BAPA \cite{absbin15} &
  \begin{tabular}[c]{@{}l@{}}Interaction fingerprint + Vina \\ energy terms\end{tabular} &
  \begin{tabular}[c]{@{}l@{}}\textbf{\underline{1D CNN}} \\ \textbf{\underline{with attention}}\end{tabular} \\
  \\
ET-score \cite{absbin16} &
  \begin{tabular}[c]{@{}l@{}}Distance weighted interaction\\ fingerprint\end{tabular} &
  \textbf{\underline{ERT}} \\
  \\
SMPLIP-Score \cite{absbin18} &
  \begin{tabular}[c]{@{}l@{}}Interaction fingerprint + ligand\\ fragment embeddings\end{tabular} &
  \textbf{\underline{RF}}, DNN \\
  \\
Taba \cite{absbin22} &
  \begin{tabular}[c]{@{}l@{}}Mass-spring distance weighted\\ interaction fingerprints\end{tabular} &
  \begin{tabular}[c]{@{}l@{}}LR, LAS, \\ lasso, \\ RR, \textbf{\underline{elastic net}}\(^a\)\end{tabular} \\
  \\
Zhu et al. (2020) \cite{absbin26} &
  Protein-ligand pairwise interactions &
  \textbf{\underline{FFNN}} \\
  \\
OnionNet \cite{absbin32} &
  \begin{tabular}[c]{@{}l@{}}Shell-established protein-ligand\\ interaction atom pair counts\end{tabular} &
  \textbf{\underline{2D CNN}} \\
  \\
Wojcikowski et al. (2018) \cite{absbin34} &
  PLEC fingerprint &
  LR, \textbf{\underline{RF}}, NN \\
  \\
Leidner et al. (2019) \cite{absbin36} &
  \begin{tabular}[c]{@{}l@{}}Protein residue centered interaction\\ FPs\end{tabular} &
  \textbf{\underline{XGBoost}} \\
  \\
PotentialNet \cite{absbin40} &
  Adjacency-based atomic interactions &
  \textbf{\underline{2D CNN}} \\
  \\
3D-RISM-AI \cite{absbin3} &
  \begin{tabular}[c]{@{}l@{}}Hydration free energy\\ properties + SASA + \\ rotatable bonds\end{tabular} &
  \begin{tabular}[c]{@{}l@{}}RR, SVM, RF,\\ \textbf{\underline{XGBoost}}\end{tabular} \\
  \\
\(\Delta_{vina}\)XGB \cite{absbin35} &
  Vina energy terms &
  \textbf{\underline{XGBoost}} \\
  \\
GXLE \cite{absbin10} &
  \begin{tabular}[c]{@{}l@{}}Molecular mechanics \\ energy terms + physical\\ interaction energy + empirical\\ interaction energy + ligand\\ descriptors\end{tabular} &
  \begin{tabular}[c]{@{}l@{}}LR, RR, DT, \\ \textbf{\underline{XGBoost}}, SVM,\\ RF, DNN\end{tabular} \\
  \\
Boyles et al. (2019) \cite{absbin23} &
  \begin{tabular}[c]{@{}l@{}}Structure-based energy descriptors\\ of protein-ligand complex + \\ ligand chemical descriptors\end{tabular} &
  \textbf{\underline{RF}} \\
  \\
Fujimoto et al. (2022) \cite{absbin5} &
  \begin{tabular}[c]{@{}l@{}}PMF + ligand MACCS \\ and ECFP + custom protein\\ AA count vectors\end{tabular} &
  Lasso, light GBM \\
  \\
RASPD+ \cite{absbin28} &
  Protein/ligand chemical descriptors &
  \begin{tabular}[c]{@{}l@{}}RF, SVR, DNN, \\ LR, kNN\end{tabular} \\
  \\
\begin{tabular}[c]{@{}l@{}}PerSpectML \cite{absbin7}, \\ FPRC-GBT \cite{absbin8}\end{tabular} &
  Spectral graph properties &
  \textbf{\underline{XGBoost}} \\
  \\
Nguyen et al. (2018) \cite{absbin30} &
  Spectral graph properties &
  \begin{tabular}[c]{@{}l@{}}RF, 1D CNN, \\ \textbf{\underline{ensemble}}\end{tabular} \\
  \\
AGL-Score \cite{absbin37} &
  Spectral graph properties &
  \textbf{\underline{GBT}} \\
  \\
PPS-ML \cite{absbin44} &
  Path spectral features &
  \textbf{\underline{GBT}} \\
  \\
\begin{tabular}[c]{@{}l@{}}KDeep \cite{absbin1}, DeepAtom \cite{absbin27} \\ Pafnucy \cite{absbin38}, \\ Francoeur et al. (2020) \cite{absbin48}, \\ AK-Score \cite{absbin49}\end{tabular} &
  3D voxels &
  \textbf{\underline{3D CNN}} \\
  \\
AEScore \cite{absbin11} &
  Atomic environment vector &
  \textbf{\underline{ANI NN}} \\
  \\
GAT-Score \cite{absbin14} &
  Atom and bond feature vectors &
  \textbf{\underline{GAT}} \\
  \\
ECIFGraph::HM-Holo-Apo \cite{absbin47} &
  \begin{tabular}[c]{@{}l@{}}Protein-water \& protein-ligand-water\\ graph representations\end{tabular} &
  \textbf{\underline{Graph transformer}} \\  \hline
\multicolumn{3}{c}{\textbf{Relative Binding Affinity}} \\ \hline
DeltaDelta \cite{relbin1} &
  3D Voxels &
  \textbf{\underline{2-leg 3D CNN}} \\
  \\
Gusev et al. (2023) \cite{relbin2} &
  \begin{tabular}[c]{@{}l@{}}Path-based FPs, Morgan FP, 3D\\ molecular FP, PLEC FP and combination\\ of 3D and PLEC FPs\end{tabular} &
  \begin{tabular}[c]{@{}l@{}}RF, MLP, LR,\\ kNN, SVM, GP,\\ GP with Tanimoto \\ kernel\end{tabular}
\label{tab:bindaff_overview}\\
\caption{Overview of ML methods for binding affinity predictions of different types. Underlined are the best-performing models from the tested ones. References with an asterisk use separate protein and ligand representations as input instead of protein-ligand binding complexes and references with a double asterisk use ligand-only representations. \\
Abbreviations: SASA=solvent accessible surface area, AA=amino acid, RR=ridge regression, SVM=support vector machine, RF=random forest, CNN=convolutional neural network, XGBoost=extreme gradient boosting, MACCS=molecular access systems key fingerprint, ECFP=extended connectivity fingerprint, PMF=potential of mean force, ECIF=extended connectivity interaction features, FP=fingerprint, IFP=interaction fingerprint, GBDT=gradient boosted decision trees, GAT=graph attention network, DNN=deep neural network, GBM=gradient boosted model, PLEC=protein-ligand extended connectivity, GBT=gradient boosted trees, PADIF=protein atom score contributions derived interaction fingerprint, MLP=multilayer perceptron, LR=linear regression, DT=decision tree, SVM=support vector machine, SVR=support vector regression, FFNN=feed-forward neural network, ERT=extremely randomized trees, LAS=least absolute shrinkage, GP=gaussian processes \\
\(^a\)elastic net combines lasso and ridge parametrization in linear regression}
\end{longtable}
}
\subsubsection{Binding Affinity Classifiers}\label{sec:classaff_models}

Classifier models perform classification of the input bound complexes into classes based on binding/non-binding of the ligand or a specific binding affinity range. When looking at the type of machine learning models that exist for absolute binding affinity prediction, we can find a large range of diverse models that use similar model types as in physicochemical properties prediction but that embed the input data through different methods (Table \ref{tab:bindaff_overview}). This embedding can be performed by either using ligand-only input for single target datasets, separate protein and ligand representations, or by taking the bound protein-ligand complex and using binding energies, interaction fingerprints, or 3D representations to construct embeddings. While some other embedding and model types can be found in prediction models for absolute binding affinity, theoretically any embedding and machine learning method used for absolute binding affinity prediction can also be used for classification, with a clear example being BindScope \cite{classbin3} which is very similar to KDeep \cite{absbin1}. The main difference lies only in the final output of the models where in the case of classifiers the output represents probabilities of binding or class-specific probabilities in the case of multi-class classification.

In Morris et al. (2020) \cite{classbin5} they use only ligand information in the form of embeddings generated by a text-based transformer neural network that was pre-trained on a large number of small molecules through a translation pretraining task to predict the molecules' IUPAC names from their SMILES strings. The intermediate latent embeddings were further used with a feed-forward neural network trained to classify binders and non-binders in different single-target datasets. They showed improved performance from using such latent embeddings obtained through a pre-trained transformer network as opposed to embeddings generated by a non-pretrained model. While they report good performance across different targets, the important drawbacks of such a model are first that it can only operate on single targets, meaning that a sufficient amount of data needs to be available to train the classifier models. Second, such a method could fail to classify correctly binders and non-binders that are similar in both structural and physicochemical properties as the model does not obtain any additional target-related information.

Torng et al. (2019) \cite{classbin6} on the other hand uses separate protein and ligand representations. They do this by using graph representations of the target's binding pocket and ligand's molecular structure. These graph representations consist of 2D network graphs of either the protein binding pocket or the ligand structure. For the protein representation, nodes and edges represent residues and connections between neighboring residues respectively, while for the ligand representation, they represent the atoms and molecular bonds. Both representations pass through a GCN network and both learned latent protein pocket and ligand embeddings are concatenated before a class probability is returned in the final output layer. They further also use two encoder architectures for the protein pocket embedding which first learn latent embeddings for each protein pocket residue through neighboring residues followed by a mapping of these residue latent embeddings into a 2D feature vector. Interestingly, in this work, the use of pretraining of the protein pocket embedding layers by using an auto-encoder setup learns to recover the protein pocket network graph in an unsupervised manner. This is done as usual training sets for binding affinity classification have a limited amount of diverse targets.

Differently from the previous models, all other models and methods use representations of the bound protein-ligand complex. In vScreenML \cite{classbin1} they used Rosetta energy terms coming from the Rosetta model \cite{rosetta} for energy prediction of biological systems. These were then further coupled to an XGBoost ML model to learn non-linear relationships between the different energy terms and the classification of targets into binders and non-binders. They especially took additional care to prepare debiased training sets by selecting decoys using the Dud-e server \cite{dude} to select decoys that would match physicochemical properties with the binders but have different structural arrangements. Then low energy docking poses were generated for each selected decoy and these were mapped to minimized crystal poses of actives to ensure a good overlap of general shape and charge distribution. While this approach would enable the selection of decoys that match the physicochemical and overall structure to the binders, it could still potentially contain bias firstly, due to the use of Dud-e, which as reported \cite{bias_dude_1,bias_dude_3,bias_dude_4} and discussed in Section \ref{sec:datasets} has its own intrinsic bias, and secondly, due to the fact that initially selected decoys are structurally not completely similar to the active molecules. It could therefore be better to use more debiased datasets such as MUV \cite{muv} or LIT-PCBA \cite{lit_pcba} as discussed in Section \ref{sec:datasets}. 

In Nogueira et al. (2019) \cite{classbin2} they used protein per atom score contributions derived interaction fingerprints (PADIF) which embed the interaction patterns between ligands and their targets. They tested these embeddings with both feed-forward neural networks and SVMs and noticed that SVM had a slightly better performance across different test sets. Interestingly in this work is that they used experimentally verified decoy molecules from assay data in the CHEMBL dataset \cite{chembl}. This helps to reduce the risk of selecting false negative decoys. Further, they also performed additional tests on inter-target selectivity where active compounds were cross-docked with other target families with assigned decoy labels for those. While this could add the risk of false negatives, statistically the chance of this happening would still be low. They found that their model had sufficient sensitivity to detect differences in the change of target which means that the model was able to learn specific protein-ligand interaction terms.

In BindScope \cite{classbin3} a similar model is used as in KDeep \cite{absbin1} for absolute binding affinity prediction. Both models employ a 3D voxelized representation of the protein-ligand bound complex. Here, a 3D representation of the protein-ligand binding complex gets broken up into voxels, which are volumetric counterparts of pixels in three dimensions. Just as different RGB channels in colored images, one can define multiple channels in the fourth dimension of these voxelized representations where each channel stores chemical-based information on each voxel. Such information can be aromaticity information, occupancy, hydrogen bond donor and acceptor properties, or electrostatic interaction properties. These properties and their ranges of influence are defined by each atom and its van der Waals radius. These 4D representations are then fed to a 3D CNN architecture that learns to extract hidden features and map them to a binary probability value that represents whether the ligand is a binder or not.

Another way how 3D molecular embeddings can be introduced is through molecular graph embeddings. In Lim et al. (2019) \cite{classbin4} they construct 3D molecular network graphs from 3D representations of the bound complex. Hereby protein-ligand complexes are modeled as 3D graphs where each node represents atoms and edges represent any type of inter-atomic interaction, either bonded or non-bonded. Each node and edge can hereby be defined through physicochemical or quantum mechanical information vectors. These representations can then be fed to a graph-based ML model which learns to map the 3D graph-based structural information to binding affinities. In Lim et al. (2019) \cite{classbin4} they define only the nodes by using information on the atom's element, degree, number of attached hydrogens, valence electrons, and whether it is aromatic. Further, the atom connectivity and inter-atomic distance information is added to model both bonded and non-bonded interatomic interactions and subtractions between both adjacency matrices are performed so that the model can learn differences between both interaction types.

\subsubsection{Absolute Binding Affinity Prediction}\label{sec:abbs_aff_models}

In absolute binding prediction, the binding affinity of a small molecule to its target is predicted as a single value. The binding affinity can hereby be expressed with different metrics like the dissociation constant Kd, the inhibition constant Ki, the inhibition \(IC_{50}\) or response concentrations \(EC_{50}\) of a target at half-maximal concentration or as a more physics-based metric like the difference in Gibbs free energy \(\Delta G\) between the bound and unbound state of a target. Both Kd and Ki metrics as well as the \(\Delta G\) are complex intrinsic measures of binding affinity, meaning that they depend only on the target, ligand, and the interactions that both form. Thermodynamically, they are a result of changes in enthalpic and entropic contributions upon binding according to the formula \[\Delta G = \Delta H - T \Delta S\]. As binding affinity is an equilibrium process between the unbound and bound states of the ligand to its target, the Kd and Ki constants represent ratios between the Kon and Koff reaction constants which in their turn are ratios between the concentrations of free target and ligand and the bound complex of the two. One can convert \(\Delta G\) values to Kd/Ki values via \(\Delta G = RTlnKd\). \(IC_{50}\) and \(EC_{50}\) metrics on the other hand depend on the concentrations used by the target and the ligand during the assays. Therefore these depend strongly on the experimental conditions and can produce different results for the same ligands and targets when assay conditions change. Therefore, the use of these metrics in ML should be done with care ensuring similar assay conditions when obtaining the experimental binding affinity values. Failure to do so will introduce bias into the models, harming their accuracy.

Again, diverse embedding and machine learning methods can be found. Roughly we can separate them into two main groups which were also seen previously in Section \ref{sec:classaff_models}: (1) models that use protein and ligand information where each is presented as separate structures without the explicit information of the binding complex that they form, (2) models that use protein-ligand complex conformation information. In the next sections, a more detailed overview will be given of the models in each group.

\textbf{Separate protein and ligand information.}\label{sec:sep_protlig} 
As high-quality crystallographic binding poses between a ligand and its target are expensive to obtain and docking poses rely heavily on the performance of the docking software, some models try to learn binding information from separate protein and ligand structures. In this group of models, input structures are provided as separate protein and ligand representations without any crystal or docked ligand poses into the protein's binding pocket. This allows them to learn on a larger amount of data for which high-quality binding poses are currently non-existent. 

From the models studied in this review, this is mostly done through the embedding of the ligand's SMILES representation and the protein's representation which can be generated either as a concatenation of individual ligand's embeddings for the same protein target \cite{absbin17} or through the embedding of the protein's sequence string \cite{absbin19,absbin33,absbin39}. These embeddings are usually generated through natural language processing (NLP) models which learn to embed the string representations of ligands and proteins into a 2D vector embedding. 

This embedding is further concatenated and used in various ML models ranging from simple boosted trees algorithms like XGBoost \cite{absbin17} and light gradient boosted trees \cite{absbin19} models to more advanced CNN models \cite{absbin39} which can additionally use attention weights \cite{absbin33} or long-short term memory (LSTM) blocks \cite{absbin19} to better learn long-distance relationships in these merged embedding vectors. 

Alternatively, protein and ligand information can also be presented through pairwise similarity matrices that represent how similar each protein and ligand are to other proteins and ligands in the training set. Hereby, one can employ classic distance measures like the Tanimoto similarity on 2D ligand embeddings through any embedding method such as Morgan fingerprints \cite{morgan_fp} and protein sequence similarities as computed through programs like the Basic Local Alignment Search Tool (BLAST) \cite{blast}. Such representations use the idea that similar ligands bind to similar types of protein targets and therefore would also exhibit similar binding affinity properties. These representations can then be fed to 2D CNN models that are able to extract hidden relationships between the similar ligand and target information to estimate binding affinity for new, unknown ligands or targets.

\textbf{Protein-ligand complex information.}\label{sec:protlig_complex}
While models using information from separate protein and ligand structures have shown to have a decent performance, they omit important structural information from the 3D binding complex between a ligand and its targets. Therefore, a large number of models exist that try to learn explicitly from the binding information. To do this, binding complexes are provided either as crystallographic or docked poses and embedded via various methods such as interaction fingerprints, molecular descriptors of binding, spectral graph properties, or voxel-based or graph-based representations of the binding complex. 

A largely used embedding method is the use of interaction fingerprints. These fingerprints, just as their molecular structural embedding counterparts such as Morgan \cite{morgan_fp}, the Extended Connectivity Fingerprint (ECFP) \cite{ecfp} or MACCS \cite{maccs} fingerprints used for small molecule compounds, embed specifically the interaction patterns between ligand compounds and their targets. Various such fingerprints like the Extended Connectivity Interaction Features (ECIF) \cite{absbin9} and the Protein-Ligand Extended Connectivity (PLEC) \cite{absbin34} use similar methodologies as in the classical Morgan \cite{morgan_fp}, ECFP \cite{ecfp} or MACCS \cite{maccs} fingerprints to embed information between neighboring ligand and protein groups. Hereby, ligand and protein-specific information can be represented with different types of information. For the ligand, this can be atomic element information, explicit valence information, number of bonded heavy atoms, number of bonded hydrogens, aromaticity, and ring membership. For the protein, this can be atomic element and residue information. The distance between the groups can be defined in several ways. One can incorporate the distance between central atoms of the functional groups in the ligand and protein side into the vector \cite{absbin16,absbin34}, whereby the distance is rigid, or dynamic, modeled by mass-spring-like functions \cite{absbin22}. One can also generate multiple distance shell radii \cite{absbin32} and embed interaction information for each such shell individually and concatenate all this information to obtain interaction embedded information across multiple distances for each ligand functional group or atom. 

Usually, such interaction fingerprints are ligand centric, meaning that they start from the ligand atoms or functional groups and embed both ligand and protein information in their direct neighborhood. Alternatively, one can also construct them protein-centric \cite{absbin36} whereby one would embed close ligand information around protein residues located in the binding pocket to obtain concatenated interaction information for the different protein residues. Here they would use features such as contact van der Waals potential between residue-ligand, protein-ligand hydrogen bonds, protein-ligand halogen bonds, protein-ligand salt bridges, \(\pi\)-interactions, and \(\pi\)-cation interactions. 

Such fingerprints also embed information that is in immediate proximity to the functional groups, as local protein-ligand interactions are what is driving the binding between both. One can add additional further distance information, as mentioned before, through shells \cite{absbin32} or by using all possible pairwise protein-ligand interactions \cite{absbin26}. This latter might prove to be computationally more expensive, especially with the growing size of the binding complex, and therefore more coarse-grained representations could prove to be useful there to reduce the number of pairs. In Zhu et al. (2020) \cite{absbin26} they use distance information to reduce the number of pairs and add additional quantum mechanical energy terms to the featurization such as partial charges and Lennard-Jones parameters. 

These interaction fingerprints also often use molecular structural information in the form of atomic elements, functional groups, and their bonded and non-bonded interactions. These can be further expanded to include other atomic information like formal charges, hybridization states or ring information, proteo-chemometric information \cite{absbin13}, quantum mechanical energy terms \cite{absbin15}, or ligand specific information as either chemical descriptors \cite{absbin9} or fragment embeddings \cite{absbin18}. These latter, proved \cite{absbin9} to be useful to further improve the performance of the models and in Boyles et al. (2019) \cite{absbin23} they also found that training on ligand information alone would teach the model an average binding affinity score for that ligand across its different protein targets. 

All these different interaction fingerprints are constructed as flat 2D vectors that can be fed to a wide array of ML models like linear models with or without regularizations like lasso, ridge or least absolute shrinkage, decision trees, random forests, gradient boosted trees and forests and neural networks like feed-forward neural networks or CNNs with or without additional attention mechanisms. 
Alternatively, these interaction fingerprints can also be based on adjacency matrices as in PotentialNet \cite{absbin40} taking into account adjacency information between ligand and protein atoms with additional atomic chemical and quantum mechanical descriptors. Such representation can further be fed into 2D CNN architectures to learn hidden features. Wang et al. (2021b) 
\cite{ifp_review} provides further an overview of different possible interaction fingerprints that are used for binding affinity predictions, many of which were also found in the discussed papers. Yin et al. (2023) \cite{absbin46} provides also an interesting study on how different hyperparameters that guide interaction fingerprints construction affect the performance of binding affinity prediction models.

Alternatively one can also construct flat 2D embedding vectors using quantum mechanical descriptors of protein-ligand binding interaction \cite{absbin3,absbin5,absbin10,absbin23,absbin35}. Common descriptors hereby are: hydration free energy properties, solvent-accessible surface area, information on rotatable bonds, physical and empirical interaction energies, or Vina energy terms. The Vina terms consist of protein-ligand interaction terms, ligand property counts, and buried solvent-accessible surface area features. In GXLE \cite{absbin10} they noticed that combining different energy terms usually gives better performance, especially when information embedded in these terms is highly complementary. However, in Nguyen et al. (2018) \cite{absbin30} they noticed that this is not always the case as they found out with the inclusion of additional Vina energy terms. 
These descriptor embeddings can additionally also be extended with structural ligand embeddings \cite{absbin5} using traditional embedding algorithms such as MACCS \cite{maccs} or ECFP \cite{ecfp}, chemical ligand descriptors \cite{absbin10,absbin23} or protein embedding information like amino acid count vectors \cite{absbin5}. These types of embeddings can further be fed into again a wide array of ML models like linear models, decision tree-based models, or neural networks.

Interactions can also be modeled through calculated chemical molecular descriptors, similar to the ones used in physicochemical and ADMET predictive models or in atomic embeddings in graph-based models, as seen in various other techniques described in this paper. In RASPD+ \cite{absbin28} they calculated molecular weight, number of hydrogen bond donors and acceptors, logP, molar refractivity, and the Wiener topology index for the ligands and molar refractivity, logP, hydrogen bond donor and acceptor counts and binding pocket volume for selected protein residues. The residue selection was done using various distance cutoffs depending on the type of calculated descriptor. Therefore descriptors like hydrogen bond donating and accepting groups are calculated relative to the corresponding protein or ligand structures for ligands and protein residues respectively, making them more specific towards protein-ligand interactions than their more general use in other models described in Sections \ref{sec:physchem_props} \& \ref{sec:admet}. 

Another type of flat 2D embedding is the one that uses spectral graph properties of protein-ligand binding molecular graphs. PerSpectML \cite{absbin7}, FPRC-GBT \cite{absbin8}, AGL-Score \cite{absbin37}, PPS-ML \cite{absbin44} and Nguyen et al. (2018) \cite{absbin30} use such properties to embed the protein-ligand interaction information. For this, they convert first a 3D protein-ligand complex representation into a graph where the nodes represent the atoms and edges of any form of inter-atom interactions. They use both inter-atomic distances and inter-atomic electrostatic energies between protein and ligand atoms and they specifically exclude protein-protein and ligand-ligand inter-atomic interactions. From this representation, various sub-graphs are defined based on different cutoff values of these inter-atomic distances or electrostatic energies. These sub-graphs can range from simple node cloud points to complex connected sub-graphs. For each sub-graph, graph properties, such as the sum of eigenvalues and their absolute deviation, spectral moments and spanning, are computed from their Laplacian matrices which give information on the graph connectivity. These properties at different cutoff levels are then converted to 2D flat feature vectors to be used in ML models. 

The difference in FPRC-GBT \cite{absbin8} is that while in PerSpectML \cite{absbin7} the sub-graph properties are calculated by establishing Vietoris-Rips complexes \cite{vietoris_rips}, in FPRC-GBT \cite{absbin8} this is done by establishing Ricci curvatures. Both are types of connected graphs that exhibit specific graph properties. Further, in PPS-ML \cite{absbin44} they show that this can also be done through paths that are persistent across different distance thresholds and defined on the sub-graphs Laplacian matrices. In Nguyen et al. (2018) \cite{absbin30} they apply combined approaches using different types of complexes and compute graph properties for different graph subsets that are each focused on different atom-type interactions. Also, they use an ensemble approach where graph spectral information is fed to a random forest model together with topological information through a CNN model whose outputs were further concatenated to form the final prediction. In AGL-Score \cite{absbin37} the sub-graphs are established in a similar manner as in Nguyen et al. (2018) \cite{absbin30} whereby inter-atomic interactions are modeled at different distance thresholds for different atomic element pair subsets. However, different from Nguyen et al. (2018) \cite{absbin30} is that they use simple Laplacian and adjacency matrix features calculated from their eigenvectors and values for each atomic element pair sub-graphs.

A first example of how 3D representations can directly be fed into ML models to learn their embedding and a mapping to the binding affinity property is by using 3D voxel representations of the binding complex \cite{absbin1,absbin27,absbin38,absbin48,absbin49} similarly as to BindScope \cite{classbin3} with the main difference being the output of the model where in this case it returns absolute binding affinity values. This means that the models are trained in a regression setting employing loss functions such as mean absolute error. Further, Francoeur et al. (2020) \cite{absbin48} \& AK-Score \cite{absbin49} both test ensemble models composed of multiple trained replicas of the same model architecture using different starting seeds for the weights and biases. They report improvement in performance compared with using a single model.
Another way how 3D representations can be used directly is through graph-based models as also seen in Sections \ref{sec:pka},  \ref{sec:physchem_props} \& \ref{sec:classaff_models}. Hereby, a similar embedding principle is applied as presented in Section \ref{sec:classaff_models} by featurizing nodes and edges and passing these graph representations through graph-based ML models. 

In AEScore \cite{absbin11} atomic information vectors consist of atomic environment vectors which are made up of atomic feature representations as used by the ANI model \cite{ani}, which is a neural network potential model trained to predict forces and energies of small molecules. For this, the model generates atomic features based on the atomic elements and includes radial and angular information from neighboring atoms. They further use the same architecture as the original ANI model which consists of separate neural networks for each atom type with a final pooling operation to retrieve the predicted binding affinity. Interestingly in ECIFGraph::HM-Holo-Apo \cite{absbin47} they use two input graph representations corresponding each to unbound protein-water interaction networks and protein-ligand-crystal water-bound interaction networks. They use statistical potentials to estimate water placement using the HydraMap tool \cite{hydramap}. This way, additional information on desolvation and water replacement effects can be incorporated together with crystal water-bound mediated interaction information. Such desolvation effects are important since they can infer entropic contributions to the binding, which, using previously described methods, is not possible since only a single static protein-ligand pose is used whereas entropic contributions can only be inferred from dynamic features that represent protein, ligand and solvent movements during binding. Still, a drawback to this method is that both the protein and ligand are kept static. Therefore, the dynamic information of protein and ligand conformational changes upon binding gets lost. Possible ways to include them are through trace atomic information calculated from molecular dynamics simulations, such as in the new MISATO dataset \cite{misato}, energy differences between unbound and bound states \cite{absbin35}, or through augmentation of the static single binding poses using molecular dynamics simulations \cite{md_for_binaff}.

\textbf{Performance comparison.}\label{sec:performance_absbin}
Different standard benchmark test sets, like the Comparative Assessment of Scoring Functions (CASF) versions 2007 \cite{casf2007}, 2013 \cite{casf13_1,casf13_2} and 2016 \cite{casf16}, the Astex diverse set \cite{astex} or the Community Structure-Activity Resource (CSAR) test sets \cite{csar1,csar2,csar3,csar4,csar5,csar6}, exist for absolute binding affinity prediction, which make it possible to compare different models and methodologies between each other. In Table \ref{tab:performance_absbin} we provide an overview of the best-performing models for the different methodologies discussed in Sections \ref{sec:sep_protlig} \& \ref{sec:protlig_complex} on the most used benchmark test sets. Results for other benchmarks were omitted due to a high number of missing values.

\begin{table}[H]
\centering
\resizebox{\textwidth}{!}{%
\begin{tabular}{l|lll}
\textbf{Method name} &
  \textbf{CASF07} &
  \textbf{CASF13} &
  \textbf{CASF16} \\ \hline
ECIF-LD-GBT \cite{absbin9}  &       &       & 0.866 \\
PerSpectML \cite{absbin7}  & 0.836 & 0.793 & 0.840 \\
PPS-ML \cite{absbin44} & 0.836 & 0.793 & 0.840 \\
FPRC-GBT \cite{absbin8}  & 0.831 & 0.805 & 0.834 \\
AGL-Score \cite{absbin37} & 0.830 & 0.792 & 0.833 \\
ET-score \cite{absbin16} &       &       & 0.827 \\
Boyles et al. (2019) \cite{absbin23} & 0.736 & 0.840 & 0.826 \\
KDeep \cite{absbin1}  &       &       & 0.82  \\
Wojcikowski et al. (2018) \cite{absbin34} &       & 0.77  & 0.82  \\
ECIFGraph::HM-Holo-Apo \cite{absbin47} &       &       & 0.820 \\
BAPA \cite{absbin15} &       & 0.771 & 0.819 \\
OnionNet \cite{absbin32} &       & 0.782 & 0.816 \\
AK-Score \cite{absbin49} &       &       & 0.812 \\
3D-RISM-AI \cite{absbin3}  &       &       & 0.80* \\
AEScore \cite{absbin11} &       & 0.76  & 0.80  \\
Francoeur et al. (2020) \cite{absbin48} &       &       & 0.80  \\
\(\Delta_{vina}\)XGB \cite{absbin35} &       &       & 0.796 \\
Fujimoto et al. (2022) \cite{absbin5}  &       &       & 0.79* \\
Pafnucy \cite{absbin38} &       & 0.70  & 0.78  \\
GAT-Score \cite{absbin14} &       & 0.78  & 0.776 \\
GXLE \cite{absbin10} &       &       & 0.762 \\
Zhu et al. (2020) \cite{absbin26} &       &       & 0.75  \\
PotentialNet \cite{absbin40} & 0.822 &       &       \\
SMPLIP-Score \cite{absbin18} &       & 0.771 &       \\
\end{tabular}%
}
\caption{Performance comparison of some methods on the most used benchmark test sets for absolute binding affinity prediction.
An asterisk indicates differences in the reported number of data points with the number of data points in the original test set.}
\label{tab:performance_absbin}
\end{table}

First, it is clear that an objective comparison is difficult to realize from compiled literature sources as the different models have not always been tested on the same benchmark datasets. Looking at performances on the CASF2016 benchmark test set \cite{casf16} the ECIF-LD-GBT method \cite{absbin9} could be established as the best performing model. Interestingly hereby is that ECIF-LD-GBT \cite{absbin9} is not a complex neural network model but a simple XGBoost model linked with custom-designed fingerprints incorporating interaction information and additional ligand-specific structural information. This trend has also been observed in previously described pKa and physicochemical properties prediction models (Section \ref{sec:physchem_props}). 

Further observing performances for the other methods we can see that many lay within a very small margin and therefore could be said to have very comparable performances. This means that different methodologies and combinations in terms of input structure embeddings and ML models provide very similar results and performances. On one hand, this can indicate that most of these methods embed similar types of information and learn similar data relationships in different ways. On the other hand, it can also indicate that the existing test sets like the CASF benchmark sets, are not difficult enough to highlight important differences in performance between the different methods. This could be due to the fact that the CASF benchmark test sets are constructed from random selections of the largest target clusters in the PDBBind training sets. The use of such random selection has already been brought up in various research \cite{simpd} and the dangers of overestimating the model's performances. Therefore, methods are not tested on their generalizability capability but merely on the success of training on the PDBBind training sets. It is therefore crucial to test methods on different benchmark test sets that contain unseen information than what is present in the training sets. Therefore, testing strategies as employed in Yin et al. (2023) \cite{absbin46}, ChemBoost \cite{absbin17}, DeepFusionDTA \cite{absbin19}, SimCNN-DTA \cite{absbin21}, AttentionDTA \cite{absbin33} or DeepDTA \cite{absbin39}, where testing is performed on random selections from the training set or through cross-validation, are not advisable. In general, cross-validation should be only employed during hyperparameter optimization of the ML models with final testing to be performed on held-out external test sets to avoid model construction bias towards the used test sets which can affect the model's generalizability. 

As seen from Table \ref{tab:performance_absbin}, various benchmark test sets exist with still many others, such as the benchmark test sets, like the Schrodinger benchmark set \cite{schrodinger_test}, the J\&J benchmark test set \cite{jj_test} or the Merck benchmark test set \cite{merck_test}, used to benchmark free energy perturbation (FEP) methods, or the different binding affinity prediction challenges like D3R \cite{d3r} or the Statistical Assessment of Modeling of Proteins and Ligands (SAMPL) challenges \cite{sampl}, which can provide out-of-distribution tests which align more closely to real case virtual screening scenarios in drug discovery. Hereby, it is also important to evaluate the performance of these models not only across different test sets but also across target-specific test sets as performances can change heavily on different targets as seen in SMPLIP-Score \cite{absbin18}, KDeep \cite{absbin1}, GXLE \cite{absbin10} and AEScore \cite{absbin11}. For this, the various FEP benchmark sets \cite{schrodinger_test,jj_test,merck_test} have different target-specific subsets with multiple ligands docked to the same binding pockets. Such benchmark sets can also be found in the BindingDB dataset \cite{bindingdb}. 

Alternatively is it also possible to cluster targets in other benchmark test sets like the CASF test sets \cite{casf2007,casf13_1,casf13_2,casf16} and evaluate on the highest populated target clusters. Potential disadvantages hereby can be that the targets in the clusters may not be completely the same. This could be avoided by setting higher clustering thresholds, risking hereby that the clusters may become sparsely populated. Another drawback to this method is that one can risk evaluating binding affinity predictions for the same or similar targets but on different binding pockets, which can also introduce bias and provide less accurate or less informed evaluations. 

Further, apart from choosing multiple and well-composed benchmark test sets, one must also include various evaluation metrics apart from the classical ones like mean absolute error or Pearson's correlation. These traditionally used metrics are taken over the whole test dataset and assigned equal weights to each data point. While they are good to obtain a general notion of the model's performance for the particular test set, there are other important aspects that are not reflected by these metrics. Binding affinity prediction models are normally designed to be used in further virtual screening campaigns to find potential good binding ligands out of a large set of compounds. Metrics like early enrichment factors, which focus more on the percentage of real high binders the model ranks among its top predictions, may reflect better the model's performance in such real-case scenarios. For example, in PotentialNet \cite{absbin40} \& ECIFGraph::HM-Holo-Apo \cite{absbin47} such additional metrics are used alongside the traditional ones. 

Also, training sets such as the PDBBind \cite{pdbbind} and its test sets \cite{casf2007,casf13_1,casf13_2,casf16} are constructed from high-quality crystallographic protein-ligand poses. However, during virtual screening campaigns, it is more common that compounds are docked into the target's binding pocket, as obtaining crystal poses is a time, effort and cost-expensive task and impossible to perform in a reasonable amount of time for large compound screening libraries. Therefore, binding affinity models should also be both trained and tested on re-docked or minimized binding poses as performances may drop depending on the quality of the obtained docked or minimized poses \cite{absbin18} and different docking algorithms can also affect binding affinity prediction models' performance \cite{absbin35}. 

Lastly, Li et al. (2022) \cite{contrastive_metalearning} shows additional tricks on how learning binding affinity prediction models can be improved by dividing the training dataset into smaller parts and constructing independent learners on each training subset. Each of these learners optimizes independently its hyperparameters based on combined loss functions that take into account performance within each individual subtask and also across the different sub-training instances.

\subsubsection{Relative Binding Affinity Prediction}

One of the difficulties in training ML models for binding affinity is the lack of high-quality structural data available. Observing widely used datasets for this task in Section \ref{sec:datasets} we can note that sizes do not go higher than around 20000 data points. When we look at models trained in other tasks such as image or text, we can see that datasets there can be in the millions to billions of data points. Thus, data scarcity is one of the major bottlenecks to achieving better performing models for tasks such as binding affinity. One of the ways to reduce this problem is by using data augmentation techniques where new data points are generated as task-meaningful modifications of the original input data. These modifications, when carefully chosen, can add additional information about the data to the model, improving its performance. One of the possible techniques to augment data in binding affinity predictions is by trying to predict the relative binding affinity between pairs of bound complexes.

In DeltaDelta \cite{relbin1} a 2-legged KDeep \cite{absbin1} model is used whereby each leg consists of the standard KDeep 3D voxel-based CNN model described in Section \ref{sec:protlig_complex} for the parallel embedding of both protein-ligand complexes and merging of their latent embeddings to output the prediction for the relative binding affinity. 

In Gusev et al. (2023) \cite{relbin2} they instead use 2D embedding vectors made up of path-based, Morgan \cite{morgan_fp}, 3D molecular, PLEC and combinations of 3D and PLEC fingerprints. All of these describe and embed the structural and interaction information between the protein and ligand. They also employ an automated active learning cycle coupled with quantum mechanical calculations of the \(\Delta\Delta G\) that serve as highly accurate estimations of the relative binding affinity which are then used to train ML models. During the automated cycle they also automatically search for the best model hyperparameters and model types from a selection of random forests, multilayer perceptrons, linear regression, k-nearest neighbors, SVM and Gaussian process models where the best performing model is then used to screen a large library of possible ligands, whereby the best scoring and more diverse ligands are fed to the quantum mechanical calculations for accurate validation and incorporation into the training set to improve the selected ML model. As the models do not use any multi-leg architectures like in DeltaDelta \cite{relbin1}, this setup can only be used to compute binding affinity differences between the new potential ligands and an established reference ligand whose, preferably, crystallographic pose is used for its initial embedding. This means that the best selected ML model could be different depending on the target, initial reference ligand and initial training set. Therefore it can be wise to initialize the automated active learning loop with different seeds to be sure that the optimization does not get stuck in a local optimum. As it can also be only used against a specific reference ligand, this makes the ML model less general than the one used in DeltaDelta \cite{relbin1} which can use different reference ligands and extract additional information from pairwise differences with other good and poor binding ligands with experimentally established binding affinity values.

\subsection{ADMET Properties}\label{sec:admet}

Finally, in order to obtain a successful lead molecule, it is crucial to optimize the molecules for ADMET properties. Not doing so, one risks failure in later stages of the drug development process \cite{admet_importance_dd}, losing valuable time and resources. When screening for ADMET properties, many different tests and biomarkers need to be considered that can provide insights into the different parts of ADMET and information on various ADMET assays are available in various public datasets as discussed in Section \ref{sec:datasets}. Below we first present an overview of some important properties and biomarkers used to establish ADMET endpoints followed by a discussion of different types of ML models that can be used to predict these properties.

\textbf{ADMET endpoints.}
Absorption constitutes the passage of the drug after intake into the systemic circulation. This usually can be expressed as the human oral bio-availability or the percentage of the drug that is found in the systemic circulation after intake \cite{bioavailability}. The administration can also be expressed as the area under the curve (AUC) of the plasma bio-availability of the drug \cite{auc_cmax}. Hereby the Cmax \cite{auc_cmax} or the maximal achieved concentration of the drug in the systemic circulation is important as it is then when the drug can exert its effect. Low Cmax levels can indicate reduced concentrations in the target tissues and, consequently, failure of the drug to engage with the target \cite{bioavailability}. From the other perspective, too high levels can result in toxic effects of the drug \cite{bioavailability}. One of the factors that influences the absorption of orally administered drugs besides physicochemical factors is its transit through the membrane cells of the gastro-intestinal tract. Various assays have been designed to study the transit of drugs through this membrane such as the CACO-2 model \cite{caco2}, PAMPA assay \cite{pampa} or the MDCK model \cite{mdck}. An important factor that can influence the passage through the membrane is the action of efflux transporters such as P-glycoprotein (Pgp), which are present in many human epithelial cells \cite{pgp}. These are located on the cell membrane and are responsible to pump foreign substances out of cells. 

The following important property after absorption is the distribution of the drug to the target tissues. One of the indicators that can be measured for this is the apparent volume of distribution \cite{vd}. This metric measures the degree of the drug's distribution to tissues within the body out of systemic circulation. It is calculated after intravenous injection by dividing the total amount of drug administered by the blood concentration extrapolated to time 0. The distribution of the drug towards target tissues can be hampered by non-specific binding to plasma proteins such as albumin, intracellular proteins and glycoproteins \cite{bioavailability}. Another metric important for drugs acting on the central nervous system is the permeation through the blood-brain barrier \cite{bbb}.

The metabolism of a drug is the transformation of the drug by enzymes in the body to its metabolites. This process is important for several reasons \cite{bioavailability}. First, it is an important step in the elimination of some drugs from the body by transforming them into metabolites that can be more easily excreted through the bile or urine. Second, some drugs that are precursors depend on metabolization to become active. Third, metabolism also plays an important role in toxicity as some reactive metabolites can produce adverse toxic effects. The different enzymes involved in metabolism, mostly from the CYP enzyme family \cite{cyp}, are used as bio-markers to test the drug's metabolism. Liver microsomes \cite{lm} are another important biomarker for metabolism. These are vesicles found in the endoplasmic reticulum of the hepatocyte and contain various expressed phase I and II metabolic enzymes such as CYP-enzymes, flavine monooxygenases, esterases, amidases, epoxide hydrolases and UDP glucuronyltransferases. Another important bio-marker is the induction of the Pregnane X receptor (PXR) which in turn induces the expression of genes coding for CYP enzymes, conjugation enzymes such as carboxylesterases and transporters like MDR1 \cite{pxr1,pxr2,pxr3}.

Finally, excretion is the elimination of the drug from the body \cite{bioavailability}. This happens through one of the body fluids, gases or hair, and either directly the unmodified drug is eliminated or its metabolites. Excretion is often measured as clearance \cite{bioavailability} which is the rate at which the drug is removed from the plasma divided by its plasma concentration. Total clearance \cite{tot_cl} is the combination of all the clearance pathways by which the drug can be eliminated. Linked to this is the half-life \cite{bioavailability} of the drug or the time necessary to reduce the drug's plasma concentration by half.

Toxicity reflects a drug's action that does not form part of its intended mode of action. It can be classified under several types such as \cite{tox_class}: (1) on-target toxicity involving its main target of action, (2) hypersensitivity and immune responses that occur due to interactions of the drug with targets that induce these immune reactions, (3) off-target toxicity, where the drug binds to other targets than its main intended targets, (4) bio-activation whereby the drug's metabolites interact with proteins in the body and cause toxicity through immune reactions or off-target toxicity, (5) idiosyncratic reactions that are rare, not well understood and more specific to individuals. Some targets are generally used in toxicity screening assays due to their importance such as the potassium channels encoded and regulated by the human ether-à-go-go-related gene (hERG) \cite{herg} for cardiotoxicity screening, targets involved in hepatotoxicity \cite{hepatotoxicity_biomarkers} or tests like the Ames mutagenicity test \cite{ames} that probes if the drug can cause alterations to the DNA, important for mutagenicity screening, or the micronucleus test \cite{micronucleus} for genotoxicity.

\textbf{ADMET predictive ML models.}
Different models exist for ADMET properties prediction that can be seen in Table \ref{tab:admet_overview}. Similar to models used for physicochemical properties prediction (Section \ref{sec:physchem_props}), these models also use whole molecule embeddings. For this, they can use both 2D feature vector embeddings constructed from molecular properties like in chemical embeddings and structural fingerprints or they can use molecular network graph embeddings. 

For the construction of 2D feature vectors, several chemical descriptors can be used such as molecular weight, hydrogen bond donating and accepting groups, polar surface area or drug-likeness measures \cite{admet13}, which all can be easily calculated through packages such as RDKit \cite{rdkit}. Some \cite{admet13,admet14,admet17,multitask_dnn_admet,admet1} are further extended with structural descriptors like Morgan fingerprints \cite{morgan_fp}, MACCS keys \cite{maccs}, atom-pair counts or descriptors \cite{atom_pair} or ECFPs \cite{ecfp}. Or with computed ADMET or physicochemical properties such as logP/logD, solubility, clearance, metabolic information or cellular permeability \cite{admet9,admet10,admet11,admet12,admet17}. 

Often these sets of descriptors are further reduced by removing highly correlated features, features with many missing values or that show very low variance between compounds. Specific algorithms, such as the Boruta algorithm \cite{boruta} can be additionally used to estimate feature importance and select only the most relevant features. This reduction helps to avoid unnecessary features to keep computational load efficient and can improve model performance by removing inter-correlated features which can interfere with training \cite{admet14}. 

In Orosz et al. (2022) \cite{admet14} and Doweyko (2004) \cite{3d_fp_worse} they observed an interesting drop in performance when combining multiple descriptor types such as both 2D and 3D chemical and structural descriptors. In sharp contrast, Yin et al. (2023) \cite{absbin46} found an improvement in performance for absolute binding affinity prediction models when fingerprints encoding different information types, such as structural molecular information and interaction information between targets and their ligands, were combined. This points out the importance to ensure that selected descriptors are relevant and highly informative for the prediction task and property at hand. This could potentially be evaluated through entropy-based techniques such as information gain used in decision trees. Also, an improvement in predictions was observed in Kosugi et al. (2021) \cite{admet9} by incorporating experimental results of relevant molecular properties. While one can also use computed property values, experimental results generally have a lower error and would produce more accurate embedding descriptors. 

Another group of models \cite{admet6,admet_aq_sol,lipo_admet_gnn} that use graph-based neural networks transform molecules into molecular network graphs and employ atom and bond-specific chemical descriptors, such as those described in Section \ref{sec:physchem_props}, since they model each atom and bond explicitly. 

\textbf{Performance of ADMET predictive ML models.}
Performance-wise, a wide array of ML models has been used and tested with either type of molecular embedding, ranging from simple models such as decision trees, random forests, XGBoost, SVM, kNN, MLR, MARS, partial least squares, radial basis functions, Gaussian processes to more complex neural network models like feed-forward neural networks and graph-based neural networks. In general, no clear advantage of one model over the other can be observed with performance often being very comparable between the different models \cite{admet10,admet9,admet8} and performance also being very dataset dependent. In addition, several neural network models \cite{admet1,admet3,admet8} did not seem to improve simpler models and showed overfitting for some smaller datasets \cite{admet1} in which setting, simpler models could be more beneficial. 

Some \cite{admet5,admet13,admet17} experimented with ensemble models by combining predictions from single model architectures. They reported improved performance over their single-model counterparts. AECF \cite{admet13} used hereby a genetic algorithm to select the best combinations of datapoints sampling, individual ensemble models and ensemble aggregation rules. 

Many \cite{admet2,admet3,admet14,admet6,generative_mmpa_qsar_admet,admet8,admet10,admet11,admet13,admet_aq_sol,lipo_admet_gnn,multitask_dnn_admet} constructed models for several ADMET properties. However, with the large amount of assay data available in datasets like Tox21 \cite{tox21_1,tox21_2} or CHEMBL \cite{chembl}, to the best of our knowledge, no model or application currently exists that would combine models for all possible and available ADMET properties. One possible reason could be the lack of a sufficient number of data points for some of the properties, as dataset sizes could be below 200 or even 100 data points. This is insufficient to train accurate and general models. Also, much of the ADMET data is often private and part of drug discovery projects in pharmaceutical companies \cite{admet6,generative_mmpa_qsar_admet,admet8,admet9,admet11,admet12,admet_aq_sol,lipo_admet_gnn,multitask_dnn_admet}. Their models are constructed based on project needs and possibly not all ADMET endpoints are of major importance. 

While many constructed separate models for each property, some also tested a multi-task learning approach where one single model was trained on several ADMET properties, either in parallel or sequentially. The sequential approach is usually favored when compounds in the different subsets have a low degree of overlap \cite{multitask_dnn_admet}. As this capability is native to neural network models, which are highly flexible in their architecture design, such approaches were not seen using simpler models such as random forests or SVMs. As mentioned previously in Section \ref{lipophilicity}, such a multi-task learning approach did not always perform better compared to its single-task counterpart as seen in Chemi-Net \cite{admet_aq_sol} \& Wenzel et al. (2019) \cite{multitask_dnn_admet}. In Wenzel et al. (2019) \cite{multitask_dnn_admet} they noticed that combining highly orthogonal properties does not always result in improved performance when training in a multi-task setting. This was also seen in Broccatelli et al. (2022) \cite{lipo_admet_gnn} where only complementary ADMET properties were trained together. Still, when combined well this could potentially improve training on small datasets. 

Lastly, there is also an even distribution between classification and regression models. Some models like Yuan et al. (2020) \cite{admet17} or the one from Falcòn-Cano et al. (2020a) \cite{aq_sol_adme_knime} combined classification and regression models to construct more accurate regressions that would span a smaller range of values. Hereby, classification models serve to separate compounds into one of the value ranges before generating a more accurate prediction with the regression models. In Zhou et al. (2019) \cite{admet8} such data splitting was also performed without the prior use of classification models. They also compared the regression models with their classification counterparts and observed a higher robustness of the latter under scaling of the prediction values.

\begin{longtable}{l|lllll}
\hline
\textbf{Method name} &
  \textbf{Embedding} &
  \textbf{Tested models} &
  \textbf{Properties} \\ \hline
\endfirsthead
\multicolumn{6}{c}%
{{\bfseries Table \thetable\ continued from previous page}} \\
\hline
\textbf{Method name} &
  \textbf{Embedding} &
  \textbf{Tested models} &
  \textbf{Properties} \\ \hline
\endhead
\hline
\endfoot
\endlastfoot
\multicolumn{4}{c}{\textbf{Regression  Models}} \\ \hline
\begin{tabular}[c]{@{}l@{}}Siramshetty et al. \\ (2021) \cite{admet4}\end{tabular} &
  CD &
  RF, \textbf{\underline{GCN}} &
  \begin{tabular}[c]{@{}l@{}}RLM stability assay,\\ PAMPA,\\ KAS\end{tabular} \\
  \\
\begin{tabular}[c]{@{}l@{}}MMPA-by-QSAR \\ \cite{generative_mmpa_qsar_admet}\end{tabular} &
  CD &
  \textbf{\underline{RF}} &
  lipophilicity, HLM \\
  \\
\begin{tabular}[c]{@{}l@{}}Zhu et al. (2018) \\ \cite{admet18}\end{tabular} &
  2D/3D CD &
  \begin{tabular}[c]{@{}l@{}}MLR, SVM, MARS, \\ \textbf{\underline{RF}}\end{tabular} &
  Blood-brain partitioning \\
  \\
\begin{tabular}[c]{@{}l@{}}Zhou et al. (2019) \\ \cite{admet8}\end{tabular} &
  8192-bit ECFP &
  DNN, SVM &
  \begin{tabular}[c]{@{}l@{}}HTSA solubility,\\ CYP3a4/CYP2c9/CYP2d6,\\ microsomal metabolic stability, \\ Pgp, MDCK\\ cell permeability, \\ unbound fraction\\ in microsomes/brain/plasma\end{tabular} \\
  \\
\begin{tabular}[c]{@{}l@{}}Wenzel et al. (2019) \\ \cite{multitask_dnn_admet}*\end{tabular} &
  \begin{tabular}[c]{@{}l@{}}CD + atom-pair and \\ pharmacophoric \\ donor-acceptor\\ pair desc.\end{tabular} &
  \textbf{\underline{DNN}} &
  \begin{tabular}[c]{@{}l@{}}\(Cl_met\), Caco-2,\\ metabolic liability, logD\end{tabular} \\
  \\
\begin{tabular}[c]{@{}l@{}}Kosugi et al. (2021) \\ \cite{admet9}\end{tabular} &
  \begin{tabular}[c]{@{}l@{}}CD + \\ exp. ADMET props.\end{tabular} &
  RF, GP &
  HOB \\
  \\
\begin{tabular}[c]{@{}l@{}}Obrezanova et al. \\ (2022) \cite{admet11}*\end{tabular} &
   \begin{tabular}[c]{@{}l@{}}CD + \\ exp. ADMET props.\end{tabular} &
  \begin{tabular}[c]{@{}l@{}}\textbf{\underline{GCN}}, GP, \\ XGBoost, SVM, \\ \textbf{\underline{DNN}}\end{tabular} &
  \begin{tabular}[c]{@{}l@{}}F, \(Cl_tot\), \(Vd_ss\), AUC,\\ \(C_max\), HL, CT-curves\end{tabular} \\
  \\
\begin{tabular}[c]{@{}l@{}}Kosugi et al. (2020) \\ \cite{admet12}\end{tabular} &
  \begin{tabular}[c]{@{}l@{}}CD + \\ ADMET props.\end{tabular} &
  \begin{tabular}[c]{@{}l@{}}PLS, \textbf{\underline{RBF}},\\ \textbf{\underline{RF}}, GP\end{tabular} &
  \(Cl_{tot,rat}\) \\
  \\
\begin{tabular}[c]{@{}l@{}}Yuan et al. (2020) \\ \cite{admet17}\end{tabular} &
  \begin{tabular}[c]{@{}l@{}}2D/3D CD \& SD + \\ ADMET props.\end{tabular} &
  \begin{tabular}[c]{@{}l@{}}kNN, SVM, RF, \\ boost tree, GBR, \\ \textbf{\underline{ensemble}}\end{tabular} &
  PPB \\
  \\
\begin{tabular}[c]{@{}l@{}}Miljkovic et al. \\ (2021) \cite{admet10}\end{tabular} &
  \begin{tabular}[c]{@{}l@{}}CD + \\ pred. ADMET prop. + \\ dose\end{tabular} &
  RF, XGBoost &
  \begin{tabular}[c]{@{}l@{}}AUC,\\ HOB, \(C_{max,plasma}\),\\ \(Cl_ren\), \(Cl_tot\),\\ HL, \(t_cmax\), \(Vd_{ss/IV}\)\end{tabular} \\
  \\
Chemi-Net \cite{admet_aq_sol}* &
  Atom/bond desc. &
  \textbf{\underline{GCN}} &
  \begin{tabular}[c]{@{}l@{}}AqSol, CYP3a4,\\ HLM, HOB,\\ PXR\end{tabular} \\
  \\
\begin{tabular}[c]{@{}l@{}}Broccatelli et al. \\ (2022) \cite{lipo_admet_gnn}*\end{tabular} &
  Atom/bond desc. &
  \begin{tabular}[c]{@{}l@{}}GCN, \textbf{\underline{GAT}}, MPNN, \\ AttentiveFP\end{tabular} &
  \begin{tabular}[c]{@{}l@{}}logD, \(Cl_{HLM/hepatocytes}\), \\ kinetic solubility \\ in phosphate buffer\end{tabular} \\
  \\
\begin{tabular}[c]{@{}l@{}}Lim et al. (2022) \\ \cite{admet6}\end{tabular} &
  \begin{tabular}[c]{@{}l@{}}atom/ bond descr.,\\ QM9 pred. props.,\\ CD,\\ ANI-2x energies\end{tabular} &
  \textbf{\underline{GCN}} &
  \begin{tabular}[c]{@{}l@{}}rat hepatocyte, \\ rat and human \\ microsome, rat \(Cl_{tot}\),\\ rat and human Pgp\end{tabular} \\ \hline
\multicolumn{4}{c}{\textbf{Classification  Models}} \\ \hline
Li et al. (2023) \cite{admet2} &
  CD &
  \begin{tabular}[c]{@{}l@{}}\textbf{\underline{LXGBoost}},\\ PLS\\ DA,\\ AdaBoost\end{tabular} &
  \begin{tabular}[c]{@{}l@{}}caco-2,\\ CYP3a4,\\ hERG,\\ HOB,\\ Micronucleus test\end{tabular} \\
  \\
ABERT \cite{admet3} &
  CD &
  \begin{tabular}[c]{@{}l@{}}\textbf{\underline{ABERT}}, DT,\\ RF, ERT,\\ feed forward NN,\\ RESNET\end{tabular} &
  \begin{tabular}[c]{@{}l@{}}caco-2,\\ HOB,\\ CYP3a4\end{tabular} \\
  \\
\begin{tabular}[c]{@{}l@{}}Falcòn-Cano et al. \\ (2020b) \cite{admet5}\end{tabular} &
  CD &
  \begin{tabular}[c]{@{}l@{}}XGBoost, SVM,\\ DT, MLP, \\ naive Bayes, \\ \textbf{\underline{Ensemble}}\end{tabular} &
  HOB \\
  \\
\begin{tabular}[c]{@{}l@{}}Zhou et al. (2023) \\ \cite{admet8}\end{tabular} &
  8192-bit ECFP &
  DNN, SVM &
  \begin{tabular}[c]{@{}l@{}}HTSA solubility,\\ CYP3a4/CYP2c9/CYP2d6,\\ microsomal metabolic stability, \\ Pgp, MDCK\\ cell permeability, \\ unbound fraction\\ in microsomes/brain/plasma\end{tabular} \\
  \\
\begin{tabular}[c]{@{}l@{}}Chen et al. (2023) \\ \cite{admet1}\end{tabular} &
  CD + SD &
  \begin{tabular}[c]{@{}l@{}}kNN, \textbf{\underline{SVM}}, \\ \textbf{\underline{RF}},\\ feed-forward NN,\\ \textbf{\underline{GCN}}\end{tabular} &
  hERG \\
  \\
\begin{tabular}[c]{@{}l@{}}Orosz et al. (2022) \\ \cite{admet14}\end{tabular} &
  2D/3D CD \& SD &
  \begin{tabular}[c]{@{}l@{}}\textbf{\underline{XGBoost}}, \\ FFNN\end{tabular} &
  \begin{tabular}[c]{@{}l@{}}Ames, Pgp inhibition,\\ hERG, hepatotoxicity, BBB\\ permeability, CYP2c9\end{tabular} \\
  \\
AECF \cite{admet13} &
  \begin{tabular}[c]{@{}l@{}}CD, SD, \\ drug-likeness desc.\end{tabular} &
  \begin{tabular}[c]{@{}l@{}}DA, SVM,\\ FFNN, RF, \\ max likelihood, \\ nearest centroid, \\ kNN, \textbf{\underline{ensemble}}\end{tabular} &
  \begin{tabular}[c]{@{}l@{}}Caco-2, HIA,\\ HOB, Pgp binding type\end{tabular} \\
  \\
\begin{tabular}[c]{@{}l@{}}Yuan et al. (2020) \\ \cite{admet17}\end{tabular} &
  \begin{tabular}[c]{@{}l@{}}2D/3D CD \& SD + \\ ADMET props.\end{tabular} &
  \begin{tabular}[c]{@{}l@{}}kNN, SVM, RF, \\ boost tree, \\ GBR, \\ \textbf{\underline{ensemble}}\end{tabular} &
  PPB \\
\label{tab:admet_overview}\\
\caption{Overview of ADMET models. Methods indicated with an asterisk use multi-task learning or a combination of single and multi-task learning. \\ Abbreviations embeddings: CD=chemical descriptors, SD=structural descriptors, props.=properties, desc.=descriptors \\ Abbreviations models: kNN=k nearest neighbours, SVM=support vector machine, GCN=graph convolutional network, XGBoost=extreme gradient boosting, AdaBoost=adaptive boosting, ABERT=adaptive boosting extremely random tree, RESNET=residual network, MLP=multi layer perceptron, GAT=graph attention network, MPNN=message passing neural network, FP=fingerprint, MLR=multivariate linear regression, MARS=multivariate adaptive regression spline, RF=random forest, LXGBoost=light XGBoost, PLS=partial least squares, DA=discriminant analysis, DT=decision trees, ERT=extreme random trees, NN=neural network, DNN=deep neural network, GP=gaussian processes, RBF=radial basis functions, DA=flexible discriminant analysis, GBR=gradient boosting regression, FFNN=feed forward neural network \\ Abbreviations properties: HOB=human oral bioavailability, KAS=kinetic aquaeous solubility, HLM=human liver microsomes, AUC=area under time plasma concentration curve, HL=elimination half life, HIA=human intestinal absorption, AqSol=aquaeous solubility, PPB=plasma protein binding, \(Cl_tot\)=total clearance, \(C_{max,plasma}\)=peak plasma concentration, \(t_cmax\)=time to peak plasma concentration, \(Vd_{ss/IV}\)=volume of distribution at steady state or after IV administration, CT-curves=concentration-time curves, \(Cl_{HLM/hepatocytes}\)=clearance in human liver microsomes or hepatocytes, \(Cl_met\)=metabolic clearance}
\end{longtable}

\subsection{Understanding Predictions}

An important aspect when developing ML models is their interpretability, especially for models involved in critical decision making like those employed in drug discovery. Interpretability methods can help to understand why the model is making certain predictions for the corresponding input data, giving validity to the predictions. They can also help to uncover hidden bias or errors in the model and hence can also assist in model development and optimisation. An important aspect of these techniques is that these methods should be consistent, accurate, faithful and stable \cite{NEURIPS2020_417fbbf2} in order to provide correct interpretations. The different modeltypes described in this work all have different possible interpretability techniques and some are easier to interpret than others.

Linear models and support vector machines (SVMs) are simple models and are therefore easier to interpret than other more complex models. For linear models, standardized feature weights used to construct the linear function can be used to indicate the importance of each feature or one can observe changes in either the outcome value or the variables when changing the value of the other respectively. SVMs are slightly more complex because they transform the input feature space into higher dimensions. Platt scaling \cite{platt_scaling} can be used here to perturb each feature of the input data point and output it to probabilities of feature importance.

Decision trees, random forests and XGBoost models can usually be interpreted through techniques like Saabas. Saabas technique \cite{saabas} tries to interpret the model as a linear combination of features and the decision rules that were applied to get to the final value. This can produce feature importance plots that visualize which features have a higher weight to get to the predicted values. While the technique is easy to use, it can suffer from low consistency \cite{saabas_bad}. Another technique that can be used are Shapley values \cite{lundberg2017unified}. This method assigns importance values to each feature and can be more reliable in terms of consistency and accuracy of the explanations. Such interpretability can prove to be very useful to validate that the model captures meaningful correlations. For example in Yuan et al. (2020) \cite{admet17} it was seen that lipophilicity-based descriptors in the feature vector had the highest importance for predicting plasma protein binding while in Zhu et al. (2018) \cite{admet18} the most important features to predict blood-brain partitioning were the topological polar surface area, log octanol-water partition coefficient, van der Waals polar surface area, number of hydrogen bond donors and solvation energies. All features that clinically are also highly correlated with their prediction values.

Neural networks are larger and more complex models and are therefore harder to interpret because of the high non-linearity and the number of parameters and are often termed "black box" models. Nonetheless, various techniques exist to analyze different parts and aspects of these models \cite{nn_interpret}. The first way is to analyze directly the weights of the network. This often is difficult for large models due to the high dimension and number of hidden layers. It also does not take interactions between the hidden neurons of the network into account. Another way is by looking at the activations of the neurons and their outputs. This takes into account layer interactions but can again be difficult to interpret for large networks due to their size and high dimensionality. One way to circumvent this is by applying dimensionality reduction techniques such as t-distributed Stochastic Neighbor Embedding (t-SNE) \cite{tsne} or Uniform Manifold Approximation and Projection (UMAP) \cite{umap} to the outputs of the layer, such as in the final embedding layers, and project this lower dimensional output in two or three dimensions. These can further be combined across different data points to get a global understanding of how the model behaves on various input data \cite{activation_atlas}. Sometimes it can however be also interesting to visually see what parts of the input data contribute higher to the obtained prediction, especially for visual inputs like images or graphs. To do this, saliency maps \cite{saliency} can be constructed that consist of the original input data together with mapped importance values for each part of this input space. These importance values are obtained by backpropagating the gradients of the predicted output through the network on to the input features for different levels of modified inputs. Graph neural networks can be tricky to interpret effectively as they have connectivities both in their graph input data between the nodes as well as throughout the graph model between the neurons. Layerwise relevance propagation (LRP) together with combined graph/model walks \cite{lrp} have been shown to be able to capture and visualize this complex connectivity. Here, neuron relevances are calculated and backpropagated according to set rules using combined graph/model walks. This produces graph saliency maps highlighting the important nodes and edges that contribute the most to the prediction. This method however, can become computationally very expensive for large input graphs and models. A simpler approach for chemical graph data is the use of counterfactuals \cite{counterfactuals} where modifications to the input chemical structures are applied for which predictions are generated. These are then compared to the original input to provide an understanding of the importance of different chemical groups in the input structure for the generated predictions. On top of that it can, at the same time, produce important information for further lead optimisation of the screened compounds.

\subsection{Conclusion}
In this work we have shown the wide array of possible ML models and methods for small molecular properties predictions that can be used in drug discovery virtual screening campaigns. While the different methods use slightly different amounts and types of input information or transform them through different techniques, the reported performances lay often very close, with very limited differences in performance, which for practical applications is not too relevant. We also see that, while more complex, more flexible and computationally more expensive, neural network based models are not always able to outperform their simpler counterparts in the current context of generally low data regimes. Although, their higher flexibility can be exploited in interesting ways such as through multi-task learning in, for example, ADMET prediction models (Section \ref{sec:admet}). But, this does not always provide the better results \cite{lipo_admet_gnn}. Therefore, additional effort should be taken to validate its training strategy in order to establish best practices, and care needs to be always taken when selecting the different subtasks, as these need to be complementary to ensure successful training and gains in performance. 

While this lack of difference between the models could be attributed to the fact that all of them learn very similar relationships in the data through either explicit or implicit ways using other surrogate descriptors, it could also highlight a lack of strong and diverse benchmark test sets. While there exist various benchmark sets for the binding affinity prediction task, it was shown that many are not widely used, that there is a lack of consensus on which benchmarks to include during model testing and that some, like the CASF \cite{casf16,casf2007,casf13_1,casf13_2} test sets, resemble too closely the training data. 

For the other property prediction tasks good standard benchmarks seem to be absent or not widely used, as many report test sets taken from training data or by cross-validation performances. This, as explained earlier, can result in tests that are too representative of the training data and would therefore not provide results on the generalizability of the ML models. These issues raise a need for the establishment of better, stronger, standardized and widely accepted benchmarks. These benchmarks should provide a correct balance between in and out of domain molecules \cite{admet5} in order to test the model's generalizability without also underestimating its performance through too difficult test sets that are not reflective anymore of the real case scenarios in which these models will be used. Test sets should therefore reflect as closely as possible data found in the real case scenario applications. This can be, for example, usage of time-based splits \cite{multitask_dnn_admet,lipo_admet_gnn,admet11,admet12} instead of random compound selections as it has been shown \cite{simpd} that both random selection and held-out compound clusters both over-or underestimate the model's performance respectively. Of course, time information is not always accessible. That is why techniques such as simulated medicinal chemistry project data (SIMPD) \cite{simpd} can help to establish datasplits that are highly similar to time-based splits in drug discovery projects.

Other ways to construct test sets could also be based on the inclusion of compound-dose related combinations like in Miljkovic et al. (2021) \cite{admet10} for properties that are also dose dependant, in order to test the model's sensitivity to interpret and use such data correctly. Further, it can also help to base the compound selection for testing on scaffold clusters \cite{admet11,admet12} in order to ensure a wide, heterogenous selection of structurally different compounds for models that need to perform well on a wide variety of molecules. Initiatives such as the Huggingface platform for language models and research or the Therapeutics Data Commons \cite{tdc_1,tdc_2} in biomedical research could also help to guide the community towards this needed standardization by collecting and providing validated training and test datasets and establishing leaderboards in order to more objectively compare newly developed models and highlight differences in performance in order to drive research further faster in the right direction.  

Lastly, abundant high quality training data is also needed to be able to train high quality models. While various datasets exist (Section \ref{sec:datasets}), they often lack a sufficiently large amount of datapoints. This is especially true for certain ADMET endpoint datasets which can be just in the ranges of a couple of hundred datapoints. Certain data augmentation techniques exist however to overcome the issue such as pretraining of neural networks on general molecular structural data with further finetuning on the specific property prediction datasets, multitask learning such as in several ADMET property prediction neural network models or use of molecular dynamics simulations to augment binding affinity datasets that are comprised of 3D binding complex structures, with additional bound conformations. However, besides size, quality is another important factor as large poor quality datasets still can generate models that will underperform \cite{multitask_dnn_admet}. This can, for example, be due to class imbalances in classification models \cite{admet10,admet17}. Another problem of many biomedical datasets is the large number of missing data. This can be solved for example by labeling the data with a pretrained ML model \cite{admet11} after which these newly labelled datapoints can be incorporated into the training data. One still needs to take care hereby that the generated predictions are sufficiently trustworthy. Therefore one can use multi-fidelity models by generating predictions with multiple trained replicas of the same model and use the deviation on the different generated predictions as a metric of precision. Or one can also analyze how well the predicted data points are embedded in the training data to know if the new data is in-or out of domain.

\section{Funding}
This work was supported by the Industrial Doctorates Plan of the Secretariat of Universities and Research of the Department of Economy and Knowledge of the Generalitat of Catalonia.

\section{Declaration of Generative AI and AIassisted technologies in the writing process}
During the preparation of this work the authors used ChatGPT (OpenAI) in order to assist in the drafting of the abstract and the introductory paragraph of Section \ref{sec:datasets} on datasets. After using this tool/service, the authors reviewed and edited the content as needed
and take full responsibility for the content of the publication.

\printbibliography
\end{document}